\documentclass[a4paper,12pt]{article}
\usepackage{lineno}

\usepackage[utf8]{inputenc}
\usepackage[margin=0.9in]{geometry}
\usepackage{amssymb,amsmath}
\usepackage{graphicx}
\graphicspath{ {images/} }

\usepackage{hyperref}
\usepackage{listings}
\usepackage{algorithm}
\usepackage[noend]{algpseudocode}
\usepackage[svgnames]{xcolor}
\usepackage{authblk}
\usepackage[perpage]{footmisc}
\usepackage{setspace}
\usepackage{multirow}
\usepackage{caption}
\usepackage[toc,page]{appendix}

\usepackage{titlesec}
\usepackage{xcolor}
\usepackage{colortbl}
\usepackage{array}
\usepackage{multirow}
\newcolumntype{L}[1]{>{\raggedright\let\newline\\\arraybackslash\hspace{0pt}}m{#1}}

\providecommand{\keywords}[1]
{
  \small	
  \textbf{\textit{Keywords:}} #1
}

\usepackage{natbib}
\usepackage{graphicx}
\usepackage{setspace}

\makeatletter
\def\BState{\State\hskip-\ALG@thistlm}
\makeatother

\title{The effect of biologically mediated decay rates on modelling soil carbon sequestration in agricultural settings}

\author[1,2,3,4]{Mohammad Javad Davoudabadi\thanks{mohammadjavad.davoudabadi@hdr.qut.edu.au}}
\author[4]{Daniel Pagendam}
\author[1,2,3]{Christopher Drovandi}
\author[5]{Jeff Baldock} 
\author[1,2,3]{Gentry White}

\affil[1]{School of Mathematical Sciences, Queensland University of Technology, Australia;}
\affil[2]{Australian Research Council Centre of Excellence for Mathematical \& Statistical Frontiers (ACEMS);}
\affil[3]{QUT Centre for Data Science, Queensland University of Technology, Australia;}
\affil[4]{CSIRO Data61, GPO Box 2583, Brisbane, QLD 4001, Australia;}
\affil[5]{CSIRO Agriculture \& Food, Glen Osmond, South Australia, Australia;}

\date{}

\begin{document}

\maketitle
\doublespacing
\begin{abstract}
Microbial biomass carbon (MBC), a crucial soil labile carbon fraction, is the most active component of the soil organic carbon (SOC) that regulates bio-geochemical processes in terrestrial ecosystems. Some studies in the literature ignore the effect of microbial population growth on carbon decomposition rates. In reality, we might expect that the decomposition rate should be related to the population of microbes in the soil and have a  positive relationship with the size of the microbial biomass pool. In this study, we explore the effect of microbial population growth on the accuracy of modelling soil carbon sequestration by developing and comparing two soil carbon models that consider a carrying capacity and limit to the growth of the microbial pool. We apply our models to three datasets, two small and one large datasets, and we select the best model in terms of having the best predictive performance through two model selection methods. Through this analysis we reveal that commonly used complex soil carbon models can over-fit in the presence of both small and large time-series datasets, and our simpler model can produce more accurate predictions. We conclude that considering the microbial population growth in a soil carbon model improves the accuracy of a model in the presence of a large dataset.
\end{abstract}


\section{Introduction}
Modelling soil carbon sequestration plays a crucial role in estimating and forecasting the amount of sequestered carbon in soil and carbon emission from the soil into the atmosphere. This modelling is significant in particular in building decision support systems for land managers selling carbon credits. Selling national carbon credits is in line with the purposes of some international bodies and agreements such as the Intergovernmental Panel on Climate Change
(IPCC) and the Paris and Kyoto Protocol agreements to mitigate global warming. Computer-simulation models such as RothC \citep{jenkinson1987}, FullCAM \citep{skjemstad2004calibration}, and Century \citep{parton1988} have been developed to help make inferences about trends in carbon stocks using time series of measurements collected over many years. The core of these models is made from soil carbon components known as pools. For example, these pools in the RothC model are  decomposable plant material (DPM), resistant plant matter
(RPM), humified organic matter (HUM), microbial biomass (BIO), and inert organic matter (IOM).

Several studies attempt to quantify uncertainty in soil carbon model outputs through statistical models and sensitivity analysis (running models for different sets of parameter values) \citep{juston, paul, stamati, yeluripati2009bayesian}. Statistical SOC models have some advantages over deterministic SOC models such as RothC, and the main benefit is introducing uncertainties in an SOC model. The uncertainties could be around the parameters, model inputs, dynamics, and subsequently model predictions. Modellers attempt to improve the accuracy of soil carbon models through quantifying uncertainty in model inputs, dynamics, and uncertainties in model parameters within a framework known as Bayesian hierarchical modeling (BHM) \citep{clifford}. Complex models can lead to over-fitting when they are applied to sparse datasets to make inferences about soil carbon stocks. \citet{clifford} address the over-fitting issue by simplifying their model, neglecting some soil organic carbon (SOC) components. \citet{davoudabadi2021advanced} apply some advanced Bayesian approaches such as correlated pseudo-marginal (CPM) method and Rao-Blackwellised particle filters (RBPF) to improve the speed of computation, the accuracy, and the prediction of the model in \citet{clifford}.

Microbial biomass carbon (MBC), a crucial soil labile carbon fraction, is the most active component of the SOC that regulates bio-geochemical processes in terrestrial ecosystems \citep{paul1996soil}. MBC plays a fundamental role in the SOC dynamics and serves as a helpful indicator of changes in soil carbon stabilization and nutrient dynamics following soil management practices \citep{fierer2009global,grandy2008molecular,liang2011effects}. However, microbial processing of soil organic matter releases $CO_2$ through respiration which contributes to atmospheric $CO_2$ and thus global climate change. Some microbially-explicit SOC models are introduced in recent years by considering the effect of the MBC in SOC models \citep{luo2016toward,blagodatsky2010model, frey2013temperature, moorhead2006theoretical, riley2014long,liu2020modeling,la2019biochemical}. Also, several studies consider the complexity of the SOC model and the bio-geochemical realism of models to predict soil carbon stocks  \citep{davoudabadi2021modelling, woolf2019microbial}. In particular, \citet{davoudabadi2021modelling} compared several SOC models by developing a Bayesian model selection method (leave-future-out cross-validation (LFO-CV)) that can identify the soil carbon model with the best predictive performance in light of the available data. Two biologically realistic features are not considered in \citep{davoudabadi2021modelling}: (i) decomposition rates that are dependent on the size of the biological (microbe) carbon pool (i.e. more microbes equates to faster
rates of decomposition); and (ii) a physical ceiling on the size of the biological
(microbial) carbon pool (akin to a carrying capacity in population biology).

Some authors have considered the use of reverse Michaelis-Menten reaction kinetics in modelling soil carbon decomposition as a function of the microbial biomass in the soil \citep{woolf2019microbial, xie2020bayesian, wieder2018carbon}. Indeed, reverse Michaelis-Menten reaction kinetics is used to control the microbial decomposition of metabolic and structural litter and available soil organic matter (SOM) pools. Reverse Michaelis-Menten reaction kinetics assumes that the rate of carbon decomposition by the microbe pool, via the action of enzymes, saturates to a maximum rate as the microbial pool grows, see \citet{wieder2018carbon} for more details. This is a natural way to model the variation in the decomposition rate, with microbe populations consuming the material that they are attached to, but with each particle of substrate only supporting a limited population of microbes, the carrying capacity, and the rate of decomposition, therefore, is limited by this.

In this study, we develop two new SOC models, three and five-pool models. We modify the early version of these models, introduced by \citep{davoudabadi2021modelling} by considering a carrying capacity or upper limit on the size of the microbial pool and also allowing the size of the microbial pool to moderate the decay rates of the other pools. We call our models BIO-K models. A physical limitation in the size of the microbial biomass pool is intended to reflect that in reality, the microbial biomass pool in soil carbon models is a small proportion. However, the equations that typically govern soil carbon dynamics (e.g., RothC) do not enforce this to be the case.

The framework of our SOC models is the BHM framework used by \citet{davoudabadi2021modelling} which is a natural way to account for epistemic uncertainty (uncertainty in the bio-geochemical process dynamics) in a statistically defensible
manner. Our focus is on using these models with both the temporally sparse and large datasets. The temporally sparse datasets are two datasets from Tarlee in South Australia and Brigalow in Queensland, Australia \citep{clifford,skjemstad2004calibration}, and the large dataset is from the Rothamsted experimental station in Hertfordshire, UK \citep{rothamsted1978details}. These three sites are in different climatic regions. It shows we can apply our approaches to a variety of datasets of any climatic region, and our approaches can be successfully applied to both long-historical and also shorter, sparser datasets. We evaluate our models by two Bayesian model selection methods, leave-future-out cross-validation (LFO-CV) \citep{burkner2020approximate}, and widely applicable information criterion (WAIC) \citep{watanabe2010asymptotic}. LFO-CV is used for small and temporally sparse SOC datasets, while WAIC is more applicable for large datasets. This study is motivated by the question of whether one can gain any benefit in terms of predictive accuracy from considering biologically mediated decay rates in an SOC model. To this end, we compare the predictive accuracy of our BIO-K models with the regular three and five-pool models introduced by \citet{davoudabadi2021modelling}.

The structure of the paper is as follows. The background and description of datasets are provided in Section \ref{DataBackground}. Section \ref{Methods} is devoted to describing the model framework and methods used in this study. We present the structure of our models in Section \ref{ModelStructure}. Sections \ref{results} and \ref{SectionConclusion} present our results and  a discussion of this study.

\section{Background and Description of Datasets}\label{DataBackground}
Our model selection method is motivated by three datasets that are collected from three locations in Australia and the UK. The details of these sites are presented in the following. 
\subsection{Tarlee Dataset}
Tarlee, situated $80$ km north of Adelaide, South Australia, was an agricultural research experiment site established in $1977$ to examine the impact of management practices on agricultural productivity as a long term field experiment \citep{datasetCSIRO}. The classification of the soil of the site is a hard-setting red-brown earth with sandy loam texture. Also, the site is dominated by winter rainfall with an average of $355$ mm from April to October and has a Mediterranean climate \citep{clifford, luo, skjemstad2004calibration}. Over a 20-year period, the soil properties of the Tarlee site were monitored in three fields under different management practices. Table \ref{ManagementTreatments} in Section \ref{datasets} of the supplementary material presents the time period of management treatments that were implemented in three trial fields in Tarlee. 

\subsection{Brigalow Dataset}
Brigalow, a research station in Queensland, Australia is situated in a semi-arid, and subtropical climate, and consists of three forested catchments of 12-17 ha \citep{skjemstad2004calibration}. Within each of the catchments, three monitoring sites were established in recognition of three soil types (a duplex soil and two clays). Wheat and occasional sorghum were planted in one catchment and the other catchment was planted to buffel pasture and the last one was left under native Brigalow forest. At this site, on one catchment, continuous wheat with some sorghum was established over a 18-year period after clearing land under Brigalow (\textit{Acacia harpophylla}) in $1982$. Table \ref{BrigalowDataTable} in Section \ref{datasets} of the supplementary material shows the duration of management practices in Brigalow.

\subsection{Rothamsted Dataset}
The Broadbalk continuous wheat experiment is one of the oldest continuous agronomic experiments in the world and was conducted at Rothamsted Research, one of the oldest agricultural research institutions in the world. The Broadbalk study commenced in 1843 and wheat is grown every year on all or part of the experiment. The experiment in section 1 was divided into different plots (labelled 2 - 20) receiving different fertilizer and manure treatments each year. Some treatment plots were established by 1852. Other plots such as 2.1 (2a), 20, and 19  established or became their current size later. Table \ref{BroadbalkDataTable} in Section \ref{datasets} of the supplementary material shows the treatments and plots in Broadbalk used in this study.

\section{Model Framework and Methods}\label{Methods}
\subsection{Soil Carbon Model Framework}\label{SoilCarbonModel}

Three sources of uncertainty in a dynamical SOC model that we consider in this study are errors in the observations, randomness or uncertainty inherent in the underlying physical processes, and uncertainties in model parameters \citep{clifford}. We model these uncertainties through the observation model, the process model, and the prior which are denoted $p(\mathbf{Y} | \mathbf{X}, \boldsymbol{\theta})$, $p(\mathbf{X}| \boldsymbol{\theta})$, and $p(\boldsymbol{\theta})$, respectively. Here $\mathbf{Y}$, $\mathbf{X}$, $\boldsymbol{\theta}$, $p(.)$, and $p(.|E)$ denote observations, unobserved state process, unknown parameters, the probability density function of the enclosed random variable, and the conditional probability density function given the event $E$, respectively. The observation and process models create a model framework known as the state-space model. The state-space model describes a system using indirectly observable variables known as state (or latent) variables and observable measurement variables. Although the state variables cannot be measured directly, one can estimate unobservable state variables based on observable measurement variables that depend on the state variables \citep{andrieu2010particle, fearnhead}. These two hierarchical models typically depend on an unknown parameter vector $\boldsymbol{\theta}$. Unknown parameters are treated as random variables in the Bayesian setting and modelled through a parameter model. The model framework created by observation, process, and parameter models is known as Bayesian Hierarchical Model (BHM) that can be represented mathematically as 
\begin{align}\label{BHM hierarcy}
    p(\mathbf{Y}, \mathbf{X}, \boldsymbol{\theta}) = p(\mathbf{Y},  \mathbf{X} | \boldsymbol{\theta}) p(\boldsymbol{\theta}) = p(\mathbf{Y} | \mathbf{X}, \boldsymbol{\theta}) p(\mathbf{X} | \boldsymbol{\theta}) p(\boldsymbol{\theta});
\end{align}
\noindent where the joint distribution $p(\mathbf{Y}, \mathbf{X}, \boldsymbol{\theta})$ captures all the uncertainty in the model \citep{cressie2015statistics,berliner1996hierarchical}. Through the posterior distribution $p(\mathbf{X}, \boldsymbol{\theta}|\mathbf{Y})$ we can make inferences about soil carbon dynamics, parameters, and functions of them. This distribution can be written based on (\ref{BHM hierarcy}) as follows
\begin{align}\label{Bayes formula}
   p(\mathbf{X}, \boldsymbol{\theta}|\mathbf{Y}) = \frac{p(\mathbf{Y} | \mathbf{X}, \boldsymbol{\theta}) p(\mathbf{X} | \boldsymbol{\theta}) p(\boldsymbol{\theta})}{p(\mathbf{Y})};
\end{align}
where $p(\mathbf{Y})$ depends only on data. Evaluating the posterior may be difficult when $p(\mathbf{Y})$ is analytically intractable. Fortunately, one can make inferences through analytically intractable posterior by drawing samples from it. See \citep{allenby2006hierarchical,berliner1996hierarchical,cressie2015statistics,davoudabadi2021advanced} for more details about the state-space model and BHM.

As we apply a Bayesian approach for model fitting to quantify the uncertainty in parameters and predictions, we place a prior distribution on the unknown parameter vector $\boldsymbol{\theta}$ which is shown by $p(\boldsymbol{\theta})$ in (\ref{Bayes formula}). The prior information encompasses three categories; uninformative, weakly informative, and informative. In the case of having a small dataset or sparse dataset over time, the prior distribution becomes more influential, and informative priors can become more useful. We gain prior knowledge in this study from previous studies \citep{clifford,davoudabadi2021modelling, skjemstad2004calibration} and expert opinion. Tables \ref{TabelPrior_LogModel}, \ref{BrigalowTablePrior_LogModel}, and \ref{RothTablePrior_LogModel} (Section \ref{PriorAndProposalDists}) of the supplementary material include the model parameters and their prior probability density functions.

\subsection{Posterior Distribution Inference}\label{Methodology}
We sample from the posterior distribution $p(X_{TOC}, \boldsymbol{\theta} | \mathbf{Y})$, where $X_{TOC}$ is the mass of total SOC to estimate the changes in SOC over time and to estimate the parameters driving the sequestration of carbon. 
To do so, as the posterior distribution $p(\mathbf{X}, \boldsymbol{\theta}|\mathbf{Y})$ in (\ref{Bayes formula}) can be decomposed into two components $p(\mathbf{X}|\boldsymbol{\theta}, \mathbf{Y})p(\boldsymbol{\theta} | \mathbf{Y})$, we draw samples from that posterior distribution and preserve the components related to the SOC process $X_{TOC}$ and its parameters $\boldsymbol{\theta}$.

As the posterior distribution and the resulting likelihood are not tractable, we apply the correlated pseudo-marginal (CPM) method, one of several particle Markov chain Monte Carlo (PMCMC) methods to the model to draw samples from $p(\boldsymbol{\theta} | \mathbf{Y})$. This method improves computational efficiency with respect to other state-of-the-art PMCMC methods by correlating the estimators of the intractable likelihoods in the acceptance ratio of its algorithm. Section \ref{subsecCPM} of the supplementary material, Algorithm \ref{CPMalgorithm} provides the  CPM algorithm. Estimators of the intractable likelihoods are correlated by correlating the auxiliary random numbers used to obtain these estimators, see \citet{deligiannidis2018correlated, davoudabadi2021advanced, davoudabadi2021modelling} for more details. In this algorithm, we generate candidate parameters from appropriate proposal distributions that are presented in the supplementary material Section \ref{PriorAndProposalDists}. More precisely, proposal distributions are arbitrarily user-specified distributions. If the Markov chain is run for enough iterations, it will converge to the desired posterior distribution. A proper proposal distribution can have a significant influence on the finite-time efficiency of the Markov chain. The ideal case occurs when the proposal distribution is the desired posterior distribution which is typically unknown.

Since the SOC model is a combination of linear and non-linear sub-models, we apply the Rao-Blackwellised particle filters (RBPF) to draw samples from $p(\mathbf{X}|\boldsymbol{\theta}, \mathbf{Y})$ and estimate the marginal likelihood of the state variables. The RBPF algorithm estimates the marginal likelihood of the linear and non-linear sub-models through the Kalman Filter (KF) and bootstrap particle filter (BPF), respectively, see Algorithms \ref{euclidKF1} and \ref{euclidBF1} in Sections \ref{Sub.KF} and \ref{Sub.BPF} of the supplementary material \citep{kalman1960new, gordon1993novel, doucet2000rao, davoudabadi2021advanced}. One of the advantages of the RBPF algorithm that makes it an attractive algorithm is that it computes the exact likelihood of the linear sub-model that reduces the computational cost of the estimated likelihood dramatically.

Quantifying uncertainty of our estimate can be done in many ways, for instance, by a $95\%$ credible interval or the estimated expected value of any functions of interest. We can achieve the inference about the mass of SOC added over a period of time by the Markov chain Monte Carlo (MCMC) samples of the posterior distribution. Through MCMC samples $\lbrace (X^m,\boldsymbol{\theta} ^m) : m = 1,...,M^* \rbrace$ we represent the posterior distribution $p(\boldsymbol{X},\boldsymbol{\theta} \vert \boldsymbol{Y})$ and estimate the posterior expectation of any function $g^*(\boldsymbol{X}, \boldsymbol{\theta})$.
\begin{align}\label{MCMCestimate}
    \mathbf{E}(g^*(\boldsymbol{X}, \boldsymbol{\theta}) \vert \boldsymbol{Y}) \approx  \frac{1}{M^*} \sum_{m=1}^{M^*} g^*(X^m,\boldsymbol{\theta} ^m).
\end{align}
The accuracy of such estimates has a negligible error in the case of having a sufficiently large sample size $M^*$. In this study, $g^*(\boldsymbol{X}, \boldsymbol{\theta})$ is the change in SOC of the first year of measuring it and following year $t$. For instance, in the Tarlee dataset, it is the change in SOC to field $i$ between 1978 and the following year $t$
\begin{equation*}
    g^*(\boldsymbol{X}, \boldsymbol{\theta}) = X_{TOC(t)}^i - X_{TOC(1978)}^i;
\end{equation*}
which can be estimated through MCMC samples as shown by equation (\ref{MCMCestimate}).  $X_{TOC(t)}^i$ is the summation of estimated state variables of carbon in all pools at time $t$ and field $i$.

The Gelman and Rubin's convergence diagnostic statistic, $\hat{R}$, is a diagnostic check for assessing the quality of the MCMC samples \citep{gelman1992inference}. We check the convergence of MCMC samples to the target distribution from multiple PMCMC chains to see whether the output from each chain is indistinguishable, and this occurs when the scale reduction factor is less than 1.2 \citep{brooks1998general}.

The validity of a SOC model to establish its suitability for estimating changes in soil carbon stocks is vital, in particular, to overcome over-fitting and under-fitting problems when we estimate model parameters and conduct inference with a model. We introduce two methods for model evaluation to select between competing soil carbon models in the next section.

\subsection{Model Evaluation}\label{ModelSelectionSection}
Measuring predictive accuracy is a way to validate or compare models \citep{gelman2014understanding}. In this study, we use two approaches, leave-future-out cross-validation (LFO-CV) \citep{burkner2020approximate} and widely applicable information criterion (WAIC) \citep{watanabe2010asymptotic}, which are fully Bayesian metrics in the sense that they use the entire posterior distribution. \citet{davoudabadi2021modelling} use the LFO-CV to compare the model’s predictive accuracy for four SOC models to understand whether more complex multi-pool models offer the best predictive tool when the datasets available for inference are relatively sparse. We use the same approach for sparse datasets in this paper. To this end, we use the expected log pointwise predictive density (ELPD) as a global measure of predictive accuracy, which is
\begin{align}\label{ELPD}
    \mbox{ELPD}= \log \prod _{t=L}^{T-1}  \mathbf{E}_{\boldsymbol{\theta}|Y_{1:t}}(p(\Tilde{Y}_{t+1}|Y_{1:t},\boldsymbol{\theta})) = \sum _{t=L}^{T-1} \log \int p(\Tilde{Y}_{t+1}|Y_{1:t},\boldsymbol{\theta})p(\boldsymbol{\theta}|Y_{1:t})~d\boldsymbol{\theta};
\end{align}
where $Y_{1:T} = \{Y_1, ..., Y_T\}$ is a time series of observations, $L$ is the minimum number of observations from the series that we will require before making predictions for future data \citep{burkner2020approximate}. The parameter space of $\boldsymbol{\theta}$ in (\ref{ELPD}) includes state variables as our model is a state-space model, and we should estimate unknown state variables, see Section \ref{Pred-Density} of the supplementary material for more details. Typically, the integral in (\ref{ELPD}) is not analytically tractable, however we can estimate it through Monte-Carlo methods. This can be done by sampling $(\boldsymbol{\theta}_{1:t}^1,..., \boldsymbol{\theta}_{1:t}^S)$ from the posterior distribution $p(\boldsymbol{\theta}|Y_{1:t})$ for $t \in \{1,...,\gamma\}$ where $\gamma \in \{L,...,T-1\}$ using the particle PMCMC method described in Section \ref{Methodology} and estimate the predictive density for $\Tilde{Y}_{L+1:T}$ as follows
\begin{align}\label{PridictDensity}
    p(\Tilde{Y}_{t+1}|Y_{1:t}) \approx \frac{1}{S}\sum_{s=1}^S p(\Tilde{Y}_{t+1}|Y_{1:t}, \boldsymbol{\theta}_{1:t}^s).
\end{align}

\noindent Although the LFO-CV computes the exact ELPD (hence, exact LFO-CV), it is computationally expensive when used with a larger dataset since it requires re-running the PMCMC to process each data point in time.

The WAIC method is faster than the LFO-CV method as it requires to run the PMCMC once for all data to get the posterior distribution which makes the WAIC be suitable for larger datasets. We use the WAIC method to explore the predictive accuracy of the soil carbon dynamics models and compare the applied models to a large dataset. The WAIC value by itself is not interpretable. It could be higher than a million, close to zero, or even negative value. When comparing models fitted to the same dataset, the model with the smallest WAIC value is considered to provide the best fit to the data among the candidate models. \citet{gelman2014understanding} suggest two different ways to measure the effective number of parameters in the WAIC formula which has a general format as follows
\begin{align}\label{WAIC-Formula}
    WAIC = -2 \sum_{t = 1}^T \log \int p(Y_{t}|\boldsymbol{\theta})p(\boldsymbol{\theta}|\boldsymbol{Y})d\boldsymbol{\theta} + 2 \rho_{WAIC}.
\end{align} 
The first component of (\ref{WAIC-Formula}) is the
computed log pointwise posterior predictive density, a measure of fit, and the second component is a correction for the effective number of parameters to adjust for over-fitting. As the first component of (\ref{WAIC-Formula}) is intractable, one can estimate it using $S$ samples from the posterior
\begin{align}
    \sum_{t = 1}^T \log \int p(Y_{t}|\boldsymbol{\theta})p(\boldsymbol{\theta}|\boldsymbol{Y})d\boldsymbol{\theta} \approx \sum _{t=1}^T \log \left(\frac{1}{S}\sum_{s=1}^S p({Y}_{t}| \boldsymbol{\theta}^s)\right).
\end{align}
\citet{gelman2014understanding} recommend using $\sum _{t=1}^T Var_{post}(\log p(Y_t|\boldsymbol{\theta}))$, the posterior variance of the log predictive density, to estimate the effective number of parameters as it gives results closer to the leave-one-out cross validation (LOO-CV). The WAIC method is asymptotically equivalence with the LOO-CV as the first three terms of the Taylor expansion of WAIC match the Taylor expansion of LOO-CV and \citet{watanabe2010asymptotic} argue that, asymptotically, the latter terms have negligible contribution. We can estimate the posterior variance of the log predictive density in practice by $\sum _{t=1}^T (\frac{1}{S-1}\sum_{s=1}^S (\log p({Y}_{t}| \boldsymbol{\theta}^s) - \overline{\log p({Y}_{t}| \boldsymbol{\theta}^*)})^2)$, where $\overline{\log p({Y}_{t}| \boldsymbol{\theta}^*)}$ is the mean log probability of data point $Y_t$ across all $S$ parameter samples. WAIC implicitly assumes
that the observations are independent of each other which can be problematic with our state-space model structure as there is usually temporal dependencies in the data. A solution to overcome this is to write $\log p({Y}_{t}| \boldsymbol{\theta}^s)$ in terms of conditional distributions $\log p({Y}_{1}| \boldsymbol{\theta}^s) + \sum _{t=2}^T \log p({Y}_{t}| {Y}_{1:t-1}, \boldsymbol{\theta}^s)$ \citep{auger2020introduction}. 

As mentioned earlier, we would like to compare the performance of two SOC models in terms of gaining the best predictive tool based on aforementioned methods. The structure of the models are introduced in the next section.

\section{Model Structure}\label{ModelStructure}
\begin{figure}[ht]
\begin{center}
 \includegraphics[scale=0.5]{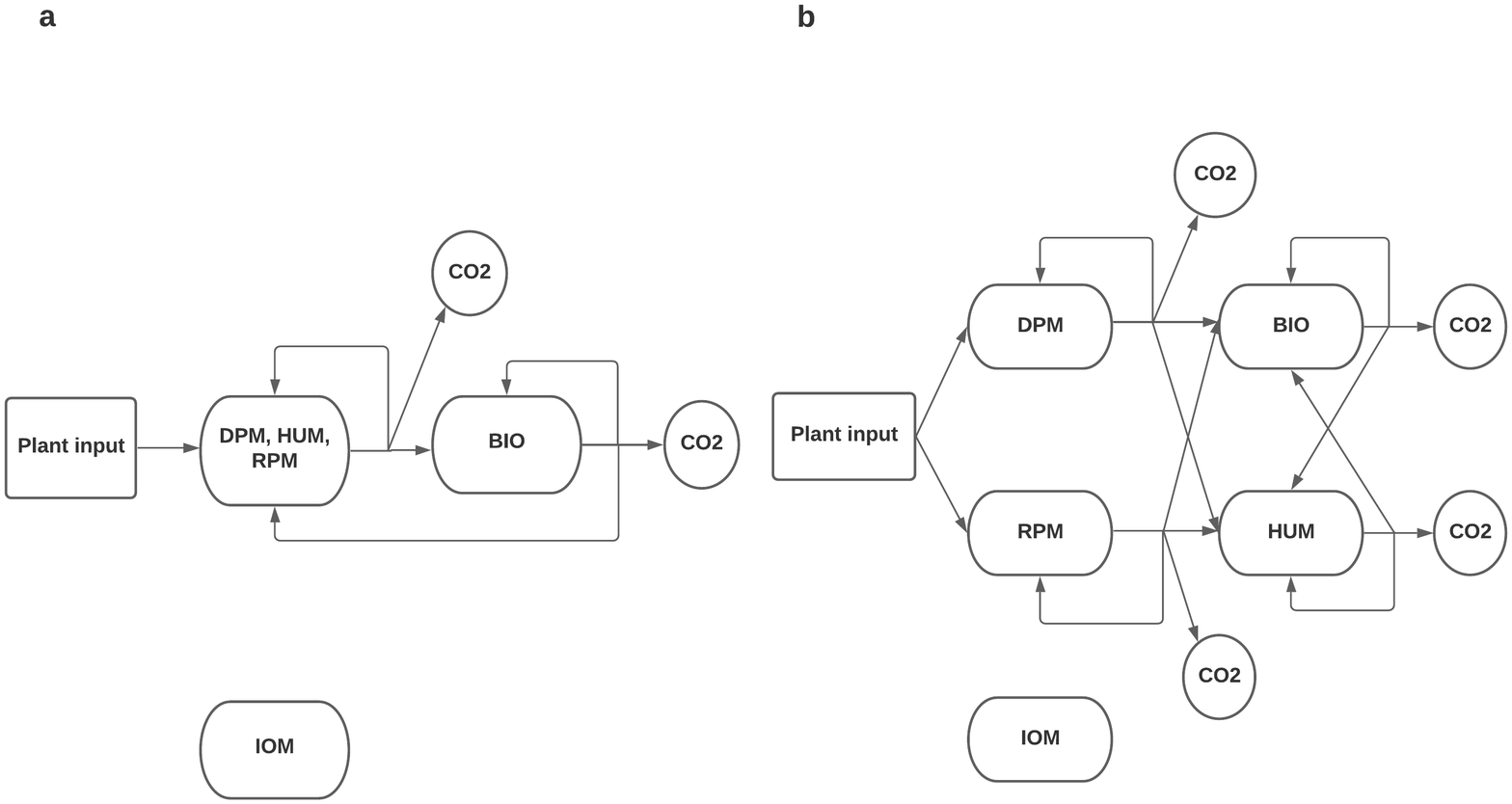}
 \vspace*{-15mm}
 \caption{Graphical representation of the carbon emission in the a) three-pool BIO-K model and b) five-pool BIO-K model. The DPM, HUM and RPM pools are amalgamated and treated as a single homogeneous pool in the three-pool model.}
\label{figThreePool}
\end{center}
\end{figure}

In this study, we introduce two new SOC models that are the modified version of the three-pool and five-pool (RothC-like) models in \citet{davoudabadi2021modelling}. We investigate how the SOC components and complexity of the SOC models affect the SOC prediction in the presence of large and small time-series datasets. The modified three-pool model consists of three conceptual pools IOM, BIO, and the main pool of decomposable carbon (an amalgamation of DPM, RPM, and HUM pools). Soil carbon decomposes from the decomposable carbon pool, and fractions are either lost to the atmosphere as $CO_2$ or transferred to the BIO pool. Carbon present in the BIO pool that decomposes is either transferred to the main soil carbon pool, lost to the atmosphere as $CO_2$, or re-assimilated as biological mass. The IOM fraction is constant since it is not subject to biological transformation. As such, the IOM process model at time t is an unknown fixed value and should be estimated. The general carbon emission in the three-pool model can also be represented graphically as depicted in Figure \ref{figThreePool}a. The process and observation models of the three-pool model are presented in detail in Sections \ref{Three_pool_process} and \ref{Three_pool_Obs} of the supplementary material, respectively. 

The general structure of the five-pool model in this study is similar to the RothC-like model introduced in \citet{davoudabadi2021modelling}. The difference is that we distribute the extra amount of carbon in the BIO pool among other pools in our model. \citet{davoudabadi2021modelling} discard the surplus amount of carbon in the BIO pool in their model. Indeed, decomposition of carbon from RPM and DPM pools either leaves the system as $CO_2$ into the atmosphere or is transformed to carbon in the HUM and BIO pools. Carbon from the HUM and BIO pools that decomposes can either be transformed to carbon in the HUM or BIO pools or lost to the atmosphere as $CO_2$.  Figure \ref{figThreePool}b shows the diagram of the carbon emission in the five-pool model. The mathematical representations of the process and observation models of the five-pool model are shown in detail in Sections \ref{Five_pool_process} and \ref{Five_pool_observation} of the supplementary material, respectively.  

\subsection{Biologically Mediated Decay Rates}
\citet{davoudabadi2021modelling} consider a constraint in their SOC pools whereby the BIO pool can be at most $5\%$ of the total SOC. This constraint prevents too much carbon from entering the microbial pool and where excess, the extra amount is rejected by rejecting BIO state trajectories in the MCMC algorithm. In reality, we might expect that the decomposition rate should be related to the population of microbes in the soil. In other words, we might expect the rates of carbon decomposition to have a positive relationship with the size of the BIO pool. In this study, we include the BIO-pool mediated decay rate in our SOC models using a logistic function with a carrying capacity, and we call our new models, in general, BIO-K models. We impose the dependency of the decomposition rate of carbon pool $p$ by 
\begin{align*}
    D_{p(t-1)} = K_p \dfrac{X_{B(t-1)}}{X_{TOC(t-1)}\kappa_{BIO}}.
\end{align*}
Where $D_{p(t-1)}$ is the decomposition rate of pool $p$ at time $t-1$, $K_p$ is the decay rate of pool $p$, $X_{B(t-1)}$ is the stock of the BIO pool at time $t-1$, $X_{TOC(t-1)}$ is the total decomposable soil carbon stock at time $t-1$, and $\kappa_{BIO} = 0.05$ is the microbial carrying capacity as a percentage of the total decomposable soil carbon. In the three-pool BIO-K model, $p$ is shown by $C$ and $B$  in the amalgamate pool (first pool) and the BIO pool, respectively, and $K_p$ is shown by $K_C$ and $K_B$ as the decay rates of the first and BIO pools, respectively. Also, in the five-pool BIO-K model, $p$ is denoted by $R$, $H$, and $B$ in pools RPM, HUM, and BIO, respectively and the associated decay rates of these pools are shown by $K_R$, $K_H$, and $K_B$, respectively.

In the three-pool BIO-K model, the extra amount of carbon in the microbial pool moves to the first pool, while in the five-pool BIO-K model this extra amount is distributed among other pools. We considered the extra amount in our models based on $U_{(t-1)}^i - \min(U_{(t-1)}^i, \kappa_{BIO}X_{Total(t-1)}^i-X_{B(t-1)}^i)$, where $U_{(t-1)}^i$ is defined for the three and five-pool BIO-K models separately in  Sections \ref{Three_pool_process} and \ref{Five_pool_process} of the supplementary material, respectively.




\section{Results}\label{results}
\subsection{Comparing Models}\label{ComparingModels}
We fit the three and five pool BIO-K models to the Brigalow and Tarlee datasets, evaluating them through the LFO-CV method as these datasets are small and temporally sparse. To do so, we worked with three PMCMC chains in the CPM method for estimating the predictive density (\ref{PridictDensity}). In both the three and five pool BIO-K models, we initialised each chain with a randomly sampled parameter vector and ran them for $200,000$ iterations discarding the first $100,000$ as burn-in in the Tarlee dataset and $400,000$ iterations discarding the first $200,000$ as burn-in in the Brigalow dataset. Since we thinned these chains, choosing every $20^{th}$ sample of the MCMC samples to estimate (\ref{PridictDensity}), $S$ in equation (\ref{PridictDensity}) for the Tarlee and Brigalow datasets are $5,000$ and $10,000$, respectively. The minimum numbers of observations, L, used for making predictions for future data were 12 and 13 in the Tarlee and Brigalow datasets, respectively. Table \ref{ELPDTandB} shows the estimated ELPD of the three and five-pool BIO-K models applied to the Tarlee and Brigalow datasets.
\begin{table}[h]
\begin{center}
\begin{tabular}{|c|l|l|l|c|}
\hline
\rowcolor[HTML]{C0C0C0} 
\cellcolor[HTML]{C0C0C0} &
  \multicolumn{2}{c|}{\cellcolor[HTML]{C0C0C0}Three-pool BIO-K model} &
  \multicolumn{2}{c|}{\cellcolor[HTML]{C0C0C0}Five-pool BIO-K model} \\ \cline{2-5} 
\rowcolor[HTML]{C0C0C0} 
\multirow{-2}{*}{\cellcolor[HTML]{C0C0C0}Dataset} &
  \multicolumn{1}{c|}{\cellcolor[HTML]{C0C0C0}Mean (ELPD)} &
  \multicolumn{1}{c|}{\cellcolor[HTML]{C0C0C0}SD} &
  \multicolumn{1}{c|}{\cellcolor[HTML]{C0C0C0}Mean (ELPD)} &
  SD \\ \hline
\cellcolor[HTML]{C0C0C0}Tarlee   & \textbf{\textcolor{red}{-35.7016}} & 1.4229 & -39.0255 & 1.8519                    \\ \hline
\cellcolor[HTML]{C0C0C0}Brigalow & \textbf{\textcolor{red}{-40.1425}} & 1.2804 & -45.9984 & 5.2612 \\ \hline
\end{tabular}
\end{center}
\caption{The mean and standard deviation (SD) of the three chains of the ELPD of the three and five-pool BIO-K models applied to the Tarlee and Brigalow datasets.}
\label{ELPDTandB}
\end{table}

\begin{table}[h]
\begin{center}
\begin{tabular}{|c|l|l|l|c|}
\hline
\rowcolor[HTML]{C0C0C0} 
\cellcolor[HTML]{C0C0C0} &
  \multicolumn{2}{c|}{\cellcolor[HTML]{C0C0C0}Three-pool model} &
  \multicolumn{2}{c|}{\cellcolor[HTML]{C0C0C0}Five-pool model} \\ \cline{2-5} 
\rowcolor[HTML]{C0C0C0} 
\multirow{-2}{*}{\cellcolor[HTML]{C0C0C0}Dataset} &
  \multicolumn{1}{c|}{\cellcolor[HTML]{C0C0C0}Mean (ELPD)} &
  \multicolumn{1}{c|}{\cellcolor[HTML]{C0C0C0}SD} &
  \multicolumn{1}{c|}{\cellcolor[HTML]{C0C0C0}Mean (ELPD)} &
  SD \\ \hline
\cellcolor[HTML]{C0C0C0}Tarlee   & \textbf{\textcolor{red}{-34.6796}} & 1.9754 & -37.2585 & 1.4591                    \\ \hline
\cellcolor[HTML]{C0C0C0}Brigalow & \textbf{\textcolor{red}{-36.2252}} & 1.7778 & -51.1718 & 4.6777 \\ \hline
\end{tabular}
\end{center}
\caption{The mean and standard deviation (SD) of the three chains of the ELPD of the three and five-pool regular models applied to the Tarlee and Brigalow datasets.}
\label{ELPDTandB_LastPaper}
\end{table}

Based on these results provided in Table \ref{ELPDTandB}, the three-pool BIO-K model outperformed the five-pool BIO-K model in the sense of gaining the best LFO predictive ability for both the Brigalow and Tarlee datasets. Table \ref{ELPDTandB_LastPaper} shows the estimated ELPD of the regular (i.e. without the additional dynamics from the BIO-K model) three and five-pool models in \citet{davoudabadi2021modelling} applied on the Tarlee and Brigalow datasets. The models in \citet{davoudabadi2021modelling} have better predictive accuracy, except the five-pool model of the Brigalow dataset, than the BIO-K models in the presence of temporally sparse datasets. In addition to that, Tables \ref{ELPDTandB} and \ref{ELPDTandB_LastPaper}  support the notion that the three-pool model has the better predictive accuracy over the five-pool model that has been frequently used for modeling soil carbon sequestration.

\begin{figure}[ht]
    \centering
    \includegraphics[width=16cm, height=10.2cm]{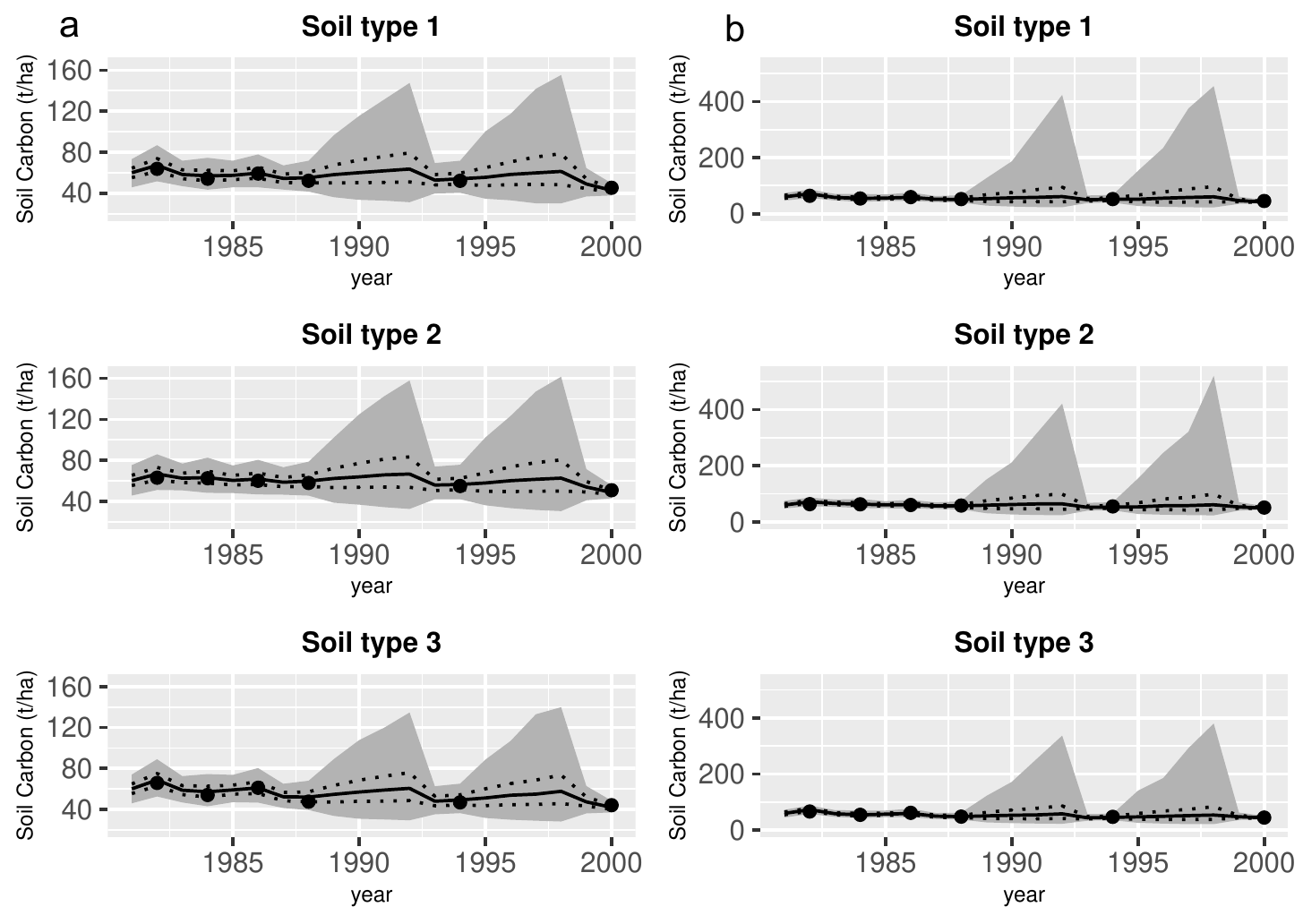}
    \vspace*{-0.5cm}
    \caption{Soil organic carbon (SOC) dynamics of the Brigalow dataset based on a) the three-pool BIO-K model and b) the five-pool BIO-K model.  The gray shaded part is the area between the $2.5^{th}$ and the $97.5^{th}$ percentiles for the SOC process gained by the three and five-pool BIO-K models. The $25^{th}$ and the $75^{th}$ percentiles for the SOC process are indicated by the dashed lines. The $50^{th}$ percentile is shown by the solid line and the measured SOC values are indicated by filled dots.}
    \label{BrigalowTrajectories}
\end{figure}

\begin{figure}[ht]
    \centering
    \includegraphics[width=16cm, height=10.2cm]{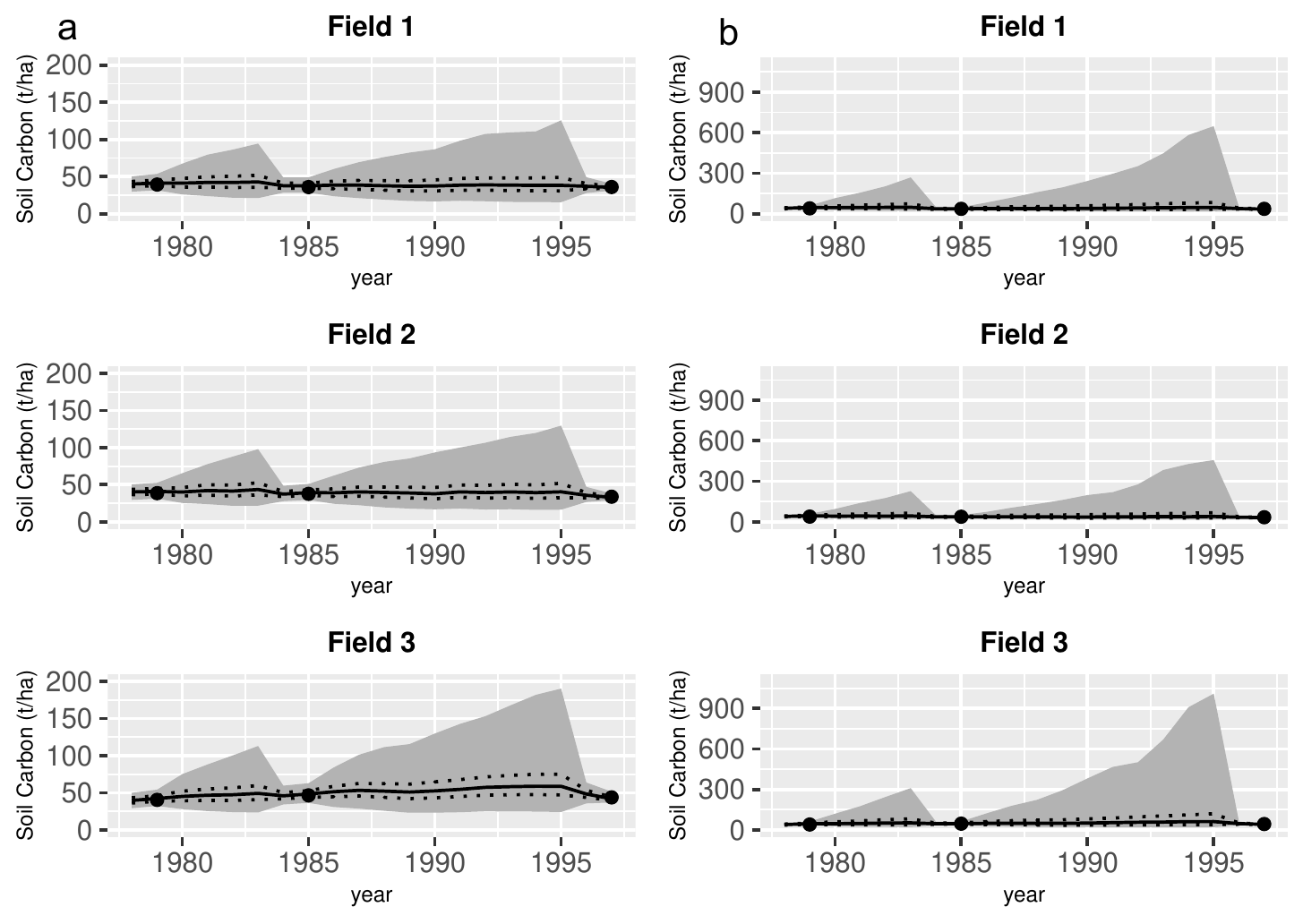}
    \vspace*{-0.5cm}
    \caption{Soil organic carbon (SOC) dynamics of the Tarlee dataset based on a) the three-pool BIO-K model and b) the five-pool BIO-K model.  The gray shaded part is the area between the $2.5^{th}$ and the $97.5^{th}$ percentiles for the SOC process gained by the three and five-pool BIO-K models. The $25^{th}$ and the $75^{th}$ percentiles for the SOC process are indicated by the dashed lines. The $50^{th}$ percentile is shown by the solid line and the measured SOC values are indicated by filled dots.}
    \label{TarleeTrajectories}
\end{figure}

Figures \ref{BrigalowTrajectories} and \ref{TarleeTrajectories} show the performance of the three and five-pool BIO-K models (columns a and b, respectively) in estimating the trajectories of the SOC dynamics of Brigalow and Tarlee datasets,  respectively. As shown in Figures \ref{BrigalowTrajectories}b and \ref{TarleeTrajectories}b, the five-pool BIO-K model increased uncertainty in the soil carbon dynamics of both datasets, especially during the sparse periods which is typified by wide $95\%$ credible intervals.

We compared the three and five-pool BIO-K models applied on the Broadbalk dataset based on the WAIC since the dataset is large. Although the SOC measurements in the Broadbalk site is sparse over time, to compare the performance of the models on large datasets, we chose plots $2b, 3, 5, 7, 10$, and $14$ as they have the most SOC measurements among other plots in the Broadbalk dataset. To compute the WAIC criterion, we worked with three PMCMC chains, each initialised with a randomly sampled parameter vector, for estimating (\ref{WAIC-Formula}). We ran each chain for 100,000 and 150,000 iterations discarding the first 20,000 and 75,000 as burn-in in the three and five-pool BIO-K models, respectively. We ran each chain for 100,000 and 200,000 iterations discarding the first 20,000 and 75,000 as burn-in for the three and five-pool regular models. Tables \ref{WAICRoth1} and \ref{WAICRoth2} show the estimated WAIC of the three and five-pool BIO-K models and regular models, respectively, applied to the Broadbalk dataset.

\begin{table}[h]
\begin{center}
\begin{tabular}{|
>{\columncolor[HTML]{C0C0C0}}c |c|l|c|l|}
\hline
\cellcolor[HTML]{C0C0C0}{\color[HTML]{333333} } &
  \multicolumn{2}{c|}{\cellcolor[HTML]{C0C0C0}Three-pool BIO-K model} &
  \multicolumn{2}{c|}{\cellcolor[HTML]{C0C0C0}Five-pool BIO-K model} \\ \cline{2-5} 
\multirow{-2}{*}{\cellcolor[HTML]{C0C0C0}{\color[HTML]{333333} MCMC Chain}} &
  \multicolumn{2}{c|}{\cellcolor[HTML]{C0C0C0}WAIC} &
  \multicolumn{2}{c|}{\cellcolor[HTML]{C0C0C0}WAIC} \\ \hline
Chain 1 & \multicolumn{2}{c|}{{\color[HTML]{FE0000} \textbf{173$\times 10^3$}}} & \multicolumn{2}{c|}{261$\times 10^3$} \\ \hline
Chain 2 & \multicolumn{2}{c|}{{\color[HTML]{FE0000} \textbf{175$\times 10^3$}}} & \multicolumn{2}{c|}{252$\times 10^3$} \\ \hline
Chain 3 & \multicolumn{2}{c|}{{\color[HTML]{FE0000} \textbf{176$\times 10^3$}}} & \multicolumn{2}{c|}{245$\times 10^3$} \\ \hline
\end{tabular}
\end{center}
\caption{The WAIC of the three MCMC chains of the three and five-pool BIO-K models applied on the Broadbalk dataset.}
\label{WAICRoth1}
\end{table}

\begin{table}[h]
\begin{center}
\begin{tabular}{|
>{\columncolor[HTML]{C0C0C0}}c |c|l|c|l|}
\hline
\cellcolor[HTML]{C0C0C0}{\color[HTML]{333333} } &
  \multicolumn{2}{c|}{\cellcolor[HTML]{C0C0C0}Three-pool regular model} &
  \multicolumn{2}{c|}{\cellcolor[HTML]{C0C0C0}Five-pool regular model} \\ \cline{2-5} 
\multirow{-2}{*}{\cellcolor[HTML]{C0C0C0}{\color[HTML]{333333} MCMC Chain}} &
  \multicolumn{2}{c|}{\cellcolor[HTML]{C0C0C0}WAIC} &
  \multicolumn{2}{c|}{\cellcolor[HTML]{C0C0C0}WAIC} \\ \hline
Chain 1 & \multicolumn{2}{c|}{{\color[HTML]{FE0000} \textbf{733$\times 10^3$}}} & \multicolumn{2}{c|}{1448$\times 10^3$} \\ \hline
Chain 2 & \multicolumn{2}{c|}{{\color[HTML]{FE0000} \textbf{732$\times 10^3$}}} & \multicolumn{2}{c|}{1533$\times 10^3$} \\ \hline
Chain 3 & \multicolumn{2}{c|}{{\color[HTML]{FE0000} \textbf{736$\times 10^3$}}} & \multicolumn{2}{c|}{1466$\times 10^3$} \\ \hline
\end{tabular}
\end{center}
\caption{The WAIC of the three MCMC chains of the three and five-pool regular models applied on the Broadbalk dataset.}
\label{WAICRoth2}
\end{table}

Based on the results shown in Tables \ref{WAICRoth1} and \ref{WAICRoth2}, the three-pool BIO-K model outperforms the five-pool BIO-K and the three-pool regular models in the sense of having lower WAIC value.

\begin{figure}[ht]
    \centering
    \includegraphics[width=16cm, height=10.2cm]{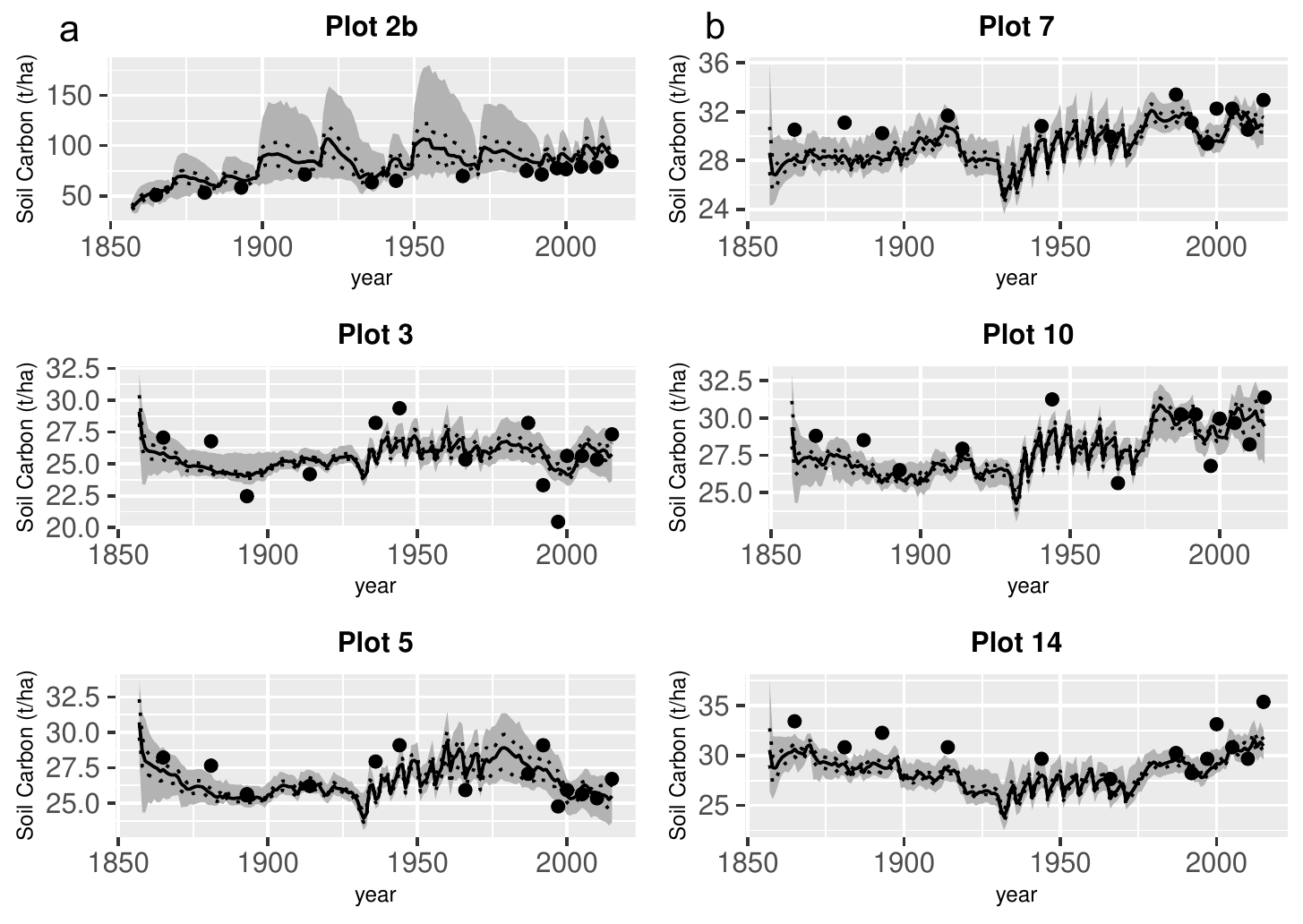}
    \vspace*{-0.5cm}
    \caption{The trajectories of the process model of the soil organic carbon (SOC) dynamics of the Broadbalk dataset based on the three-pool BIO-K model.  The gray shaded part is the area between the $2.5^{th}$ and the $97.5^{th}$ percentiles for the SOC process gained by the three-pool BIO-K model. The $25^{th}$ and the $75^{th}$ percentiles for the SOC process are indicated by the dashed lines. The $50^{th}$ percentile is shown by the solid line and the measured SOC values are indicated by filled dots.}
    \label{Roth_3P_Log_Trajectories}
\end{figure}

\begin{figure}[ht]
    \centering
    \includegraphics[width=16cm, height=10.2cm]{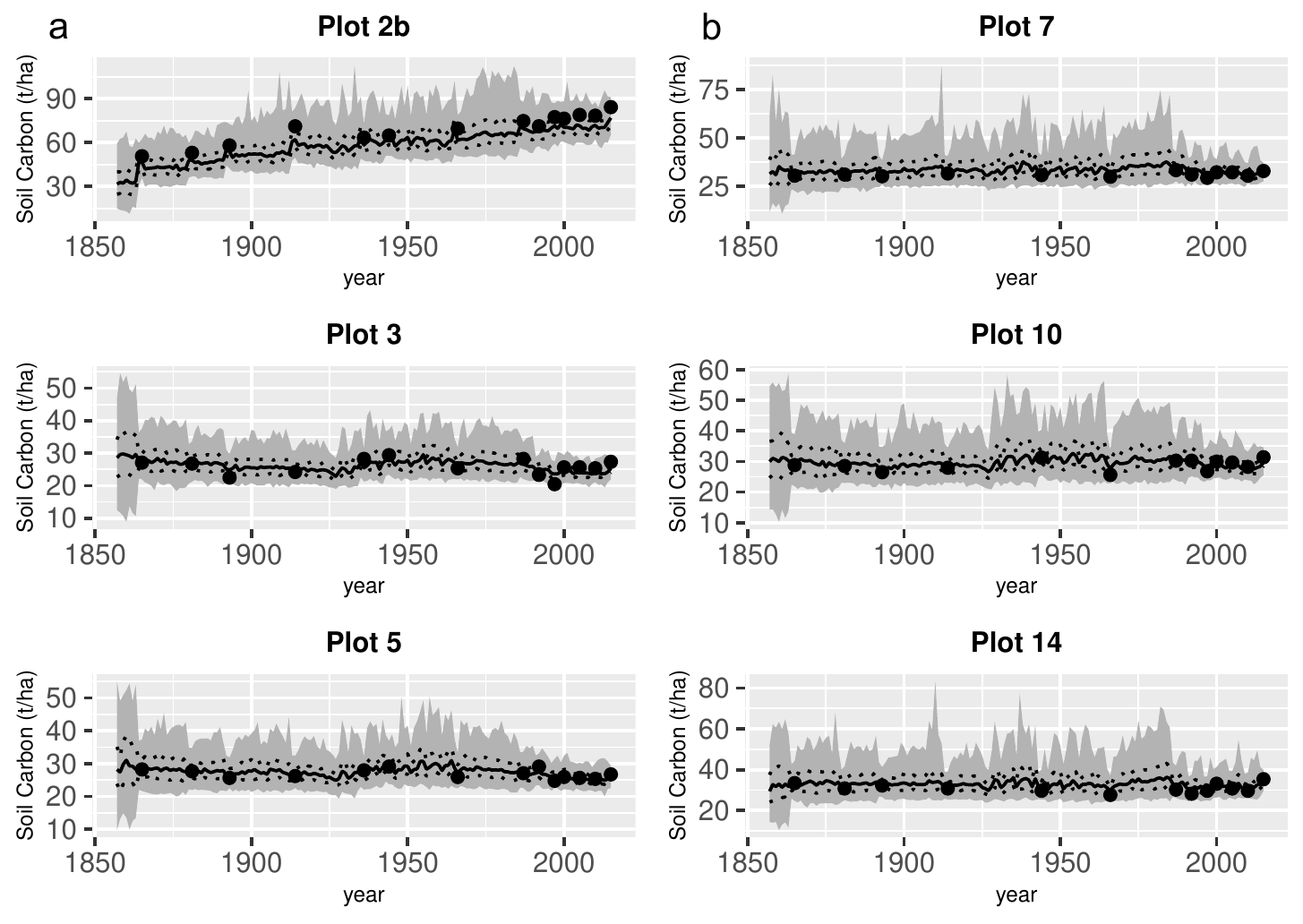}
    \vspace*{-0.5cm}
    \caption{The trajectories of the process model of the soil organic carbon (SOC) dynamics of the Broadbalk dataset based on the five-pool BIO-K model.  The gray shaded part is the area between the $2.5^{th}$ and the $97.5^{th}$ percentiles for the SOC process gained by the five-pool BIO-K model. The $25^{th}$ and the $75^{th}$ percentiles for the SOC process are indicated by the dashed lines. The $50^{th}$ percentile is shown by the solid line and the measured SOC values are indicated by filled dots.}
    \label{Roth_5P_Log_Trajectories}
\end{figure}

\begin{figure}[ht]
    \centering
    \includegraphics[width=16cm, height=10.2cm]{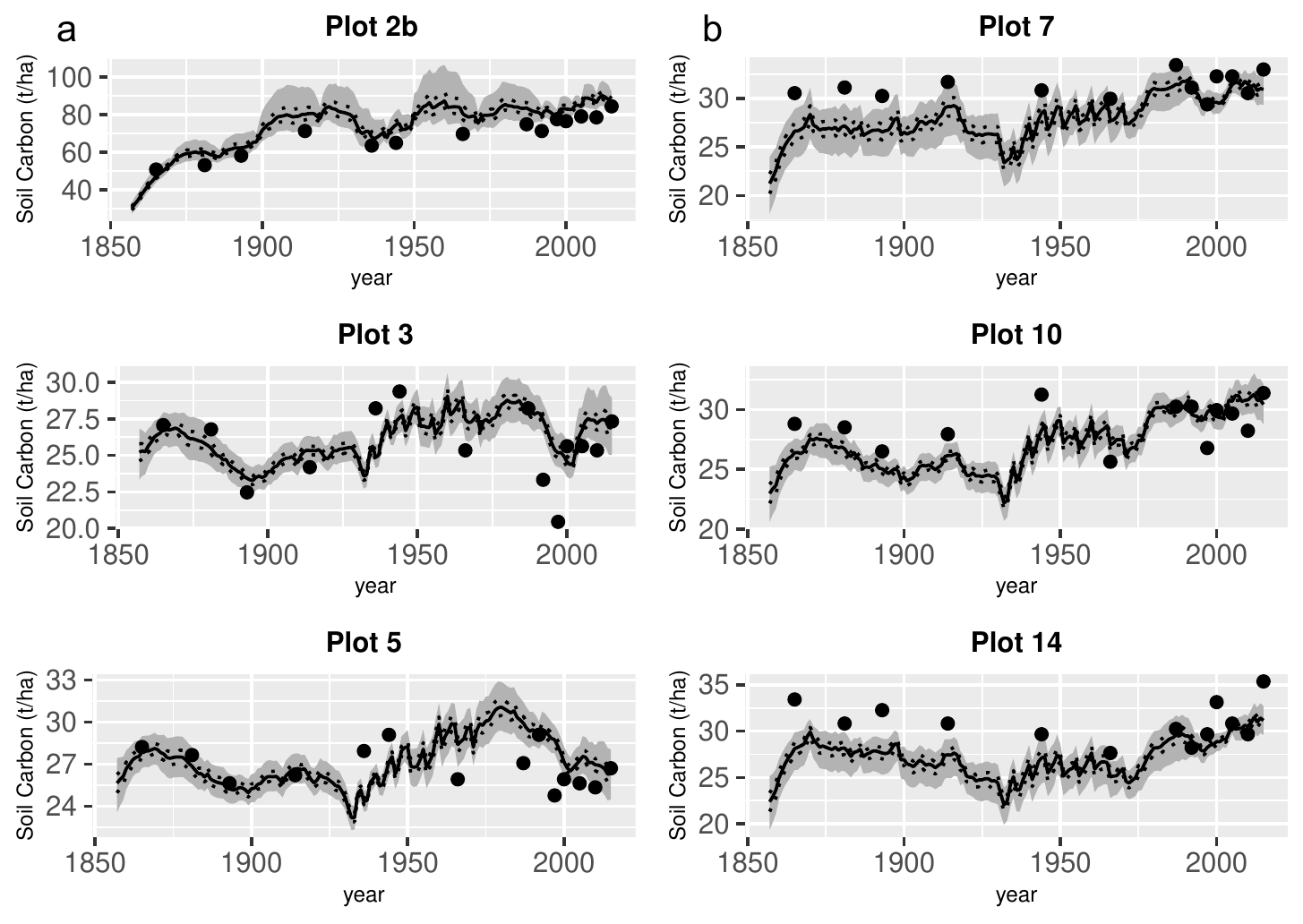}
    \vspace*{-0.5cm}
    \caption{The trajectories of the process model of the soil organic carbon (SOC) dynamics of the Broadbalk dataset based on the three-pool regular model.  The gray shaded part is the area between the $2.5^{th}$ and the $97.5^{th}$ percentiles for the SOC process gained by the three-pool regular model. The $25^{th}$ and the $75^{th}$ percentiles for the SOC process are indicated by the dashed lines. The $50^{th}$ percentile is shown by the solid line and the measured SOC values are indicated by filled dots.}
    \label{Roth_3P_NonLog_Trajectories}
\end{figure}

\begin{figure}[ht]
    \centering
    \includegraphics[width=16cm, height=10.2cm]{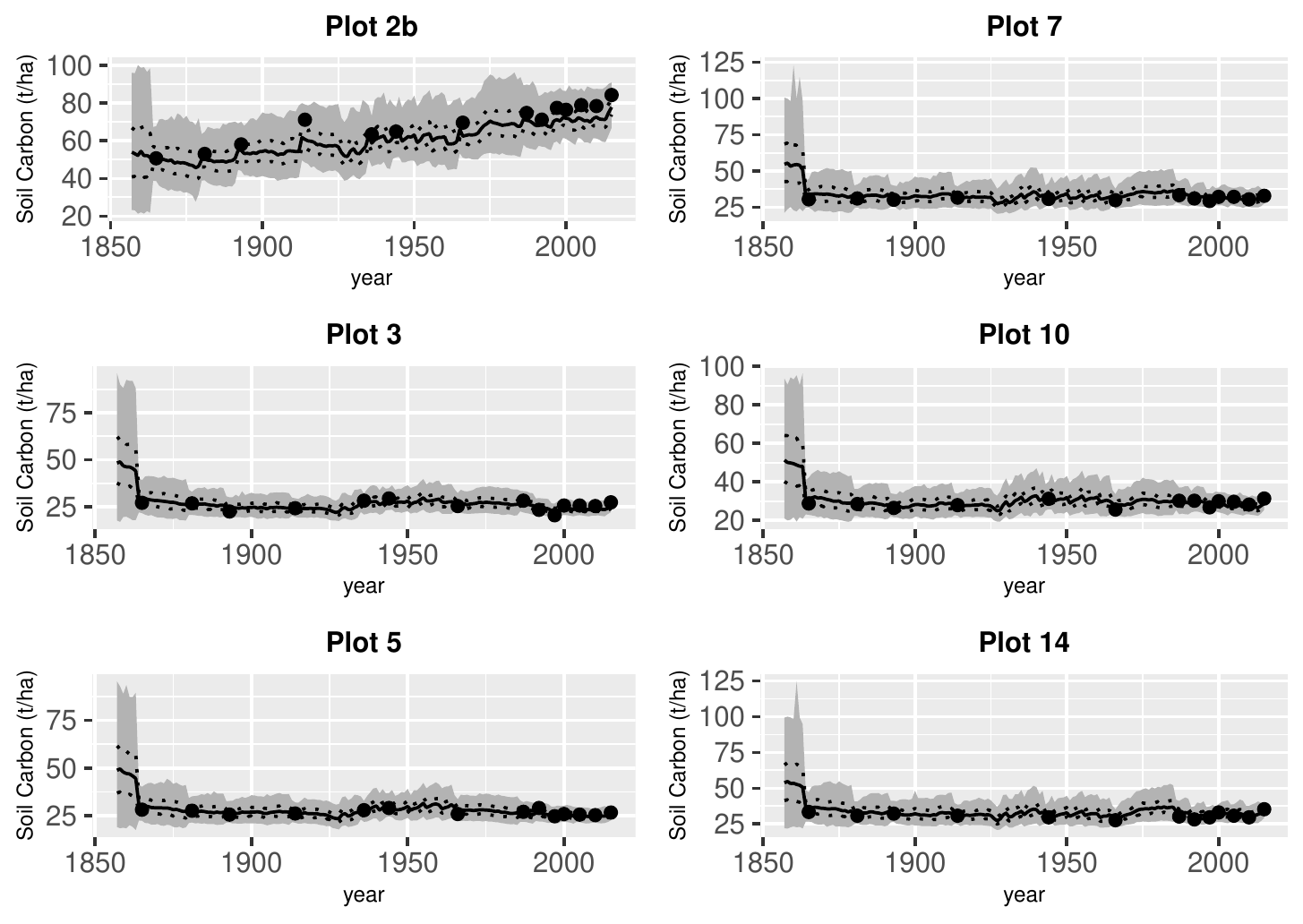}
    \vspace*{-0.5cm}
    \caption{The trajectories of the process model of the soil organic carbon (SOC) dynamics of the Broadbalk dataset based on the five-pool regular model.  The gray shaded part is the area between the $2.5^{th}$ and the $97.5^{th}$ percentiles for the SOC process gained by the five-pool regular model. The $25^{th}$ and the $75^{th}$ percentiles for the SOC process are indicated by the dashed lines. The $50^{th}$ percentile is shown by the solid line and the measured SOC values are indicated by filled dots.}
    \label{Roth_5P_NonLog_Trajectories}
\end{figure}

Figures \ref{Roth_3P_Log_Trajectories} and \ref{Roth_5P_Log_Trajectories} show the performance of the three and five-pool BIO-K models in estimating the trajectories of the SOC dynamics of the Broadbalk dataset,  respectively. The performance of the three and five-pool regular models in estimating the trajectories of the SOC dynamics of the Broadbalk dataset are shown in Figures  \ref{Roth_3P_NonLog_Trajectories} and \ref{Roth_5P_NonLog_Trajectories},  respectively. Note that the $95\%$ credible intervals are related to the trajectories of the process model while the filled dots are the observations of the latent process that have been corrupted by measurement noise.

\subsection{Uncertainty quantification} \label{inference}
Based on the three-pool BIO-K model, the average of the change in equation (\ref{MCMCestimate}) between years $1978$ and $1997$ in fields $\{1, 2, 3\}$ in Tarlee are $1.86$, $2.63$, and $14.72$, respectively. The change of SOC in the Brigalow trial between $1981$ and $2000$ base on the three-pool BIO-K model are $-0.45$, $2.93$, and $-2.70$, respectively. The average of the change of SOC in plots $\{2b,3,5,7,10,14\}$ of the Broadbalk between years $1852$ and $2015$ are $76.24$, $15.63$, $16.68$, $19.35$, $17.89$, and $18.37$, respectively. 

\begin{figure}[ht]
    \centering
    \includegraphics[width=16cm, height=10.2cm]{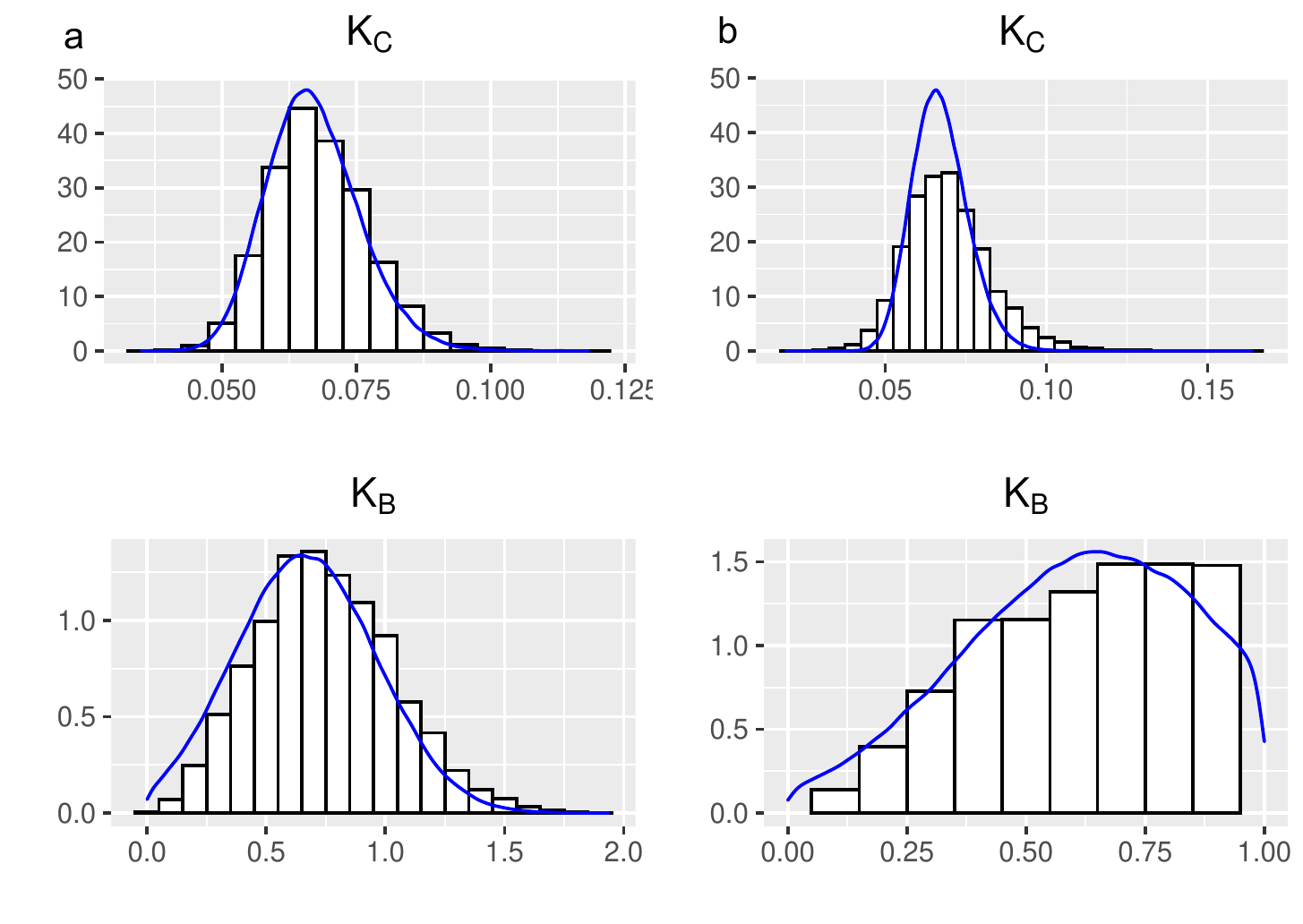}
    \vspace*{-0.5cm}
    \caption{The marginal posterior distributions (histogram) of the SOC and BIO decomposition rates, $K_C$ and $K_B$, respectively, in a) Tarlee and b) Brigalow. The histograms correspond to the three-pool BIO-K model in both Brigalow and Tarlee. The blue densities are the prior distributions of the SOC and BIO decomposition rates.}
    \label{Pri_Post_BIO-K_Tar_Brig}
\end{figure}

\begin{figure}[!ht]
    \centering
    \includegraphics[width=16cm, height=10.2cm]{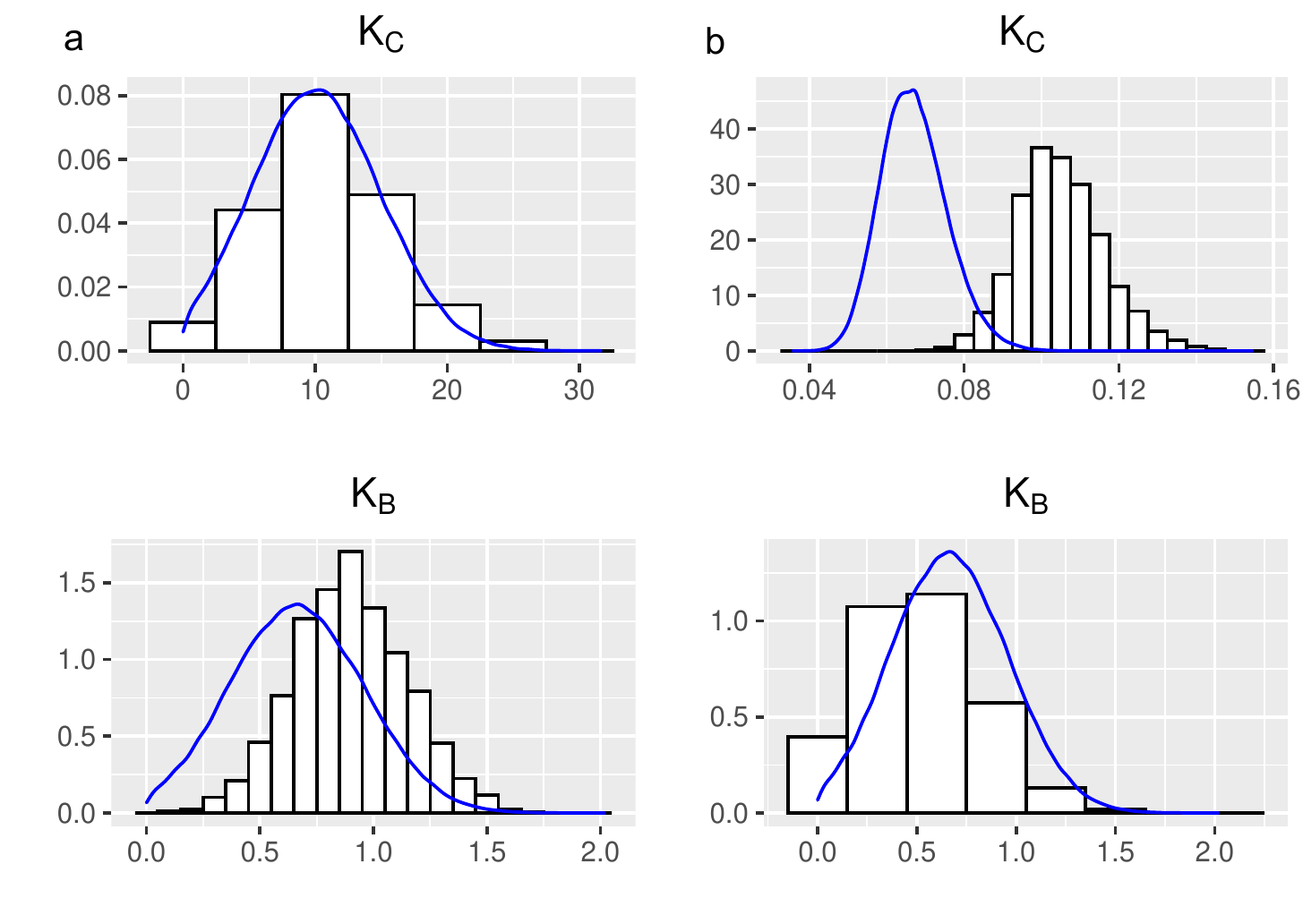}
    \vspace*{-0.5cm}
    \caption{The marginal posterior distributions (histogram) of the SOC and BIO decomposition rates, $K_C$ and $K_B$, respectively, in Broadbalk dataset. The histograms correspond to the three-pool a) BIO-K and b) regular models in Broadbalk. The blue densities are the prior distributions of the SOC and BIO decomposition rates.}
    \label{Pri_Post_3P_Roth}
\end{figure}

\begin{figure}[!ht]
    \centering
    \includegraphics[width=16cm, height=10.2cm]{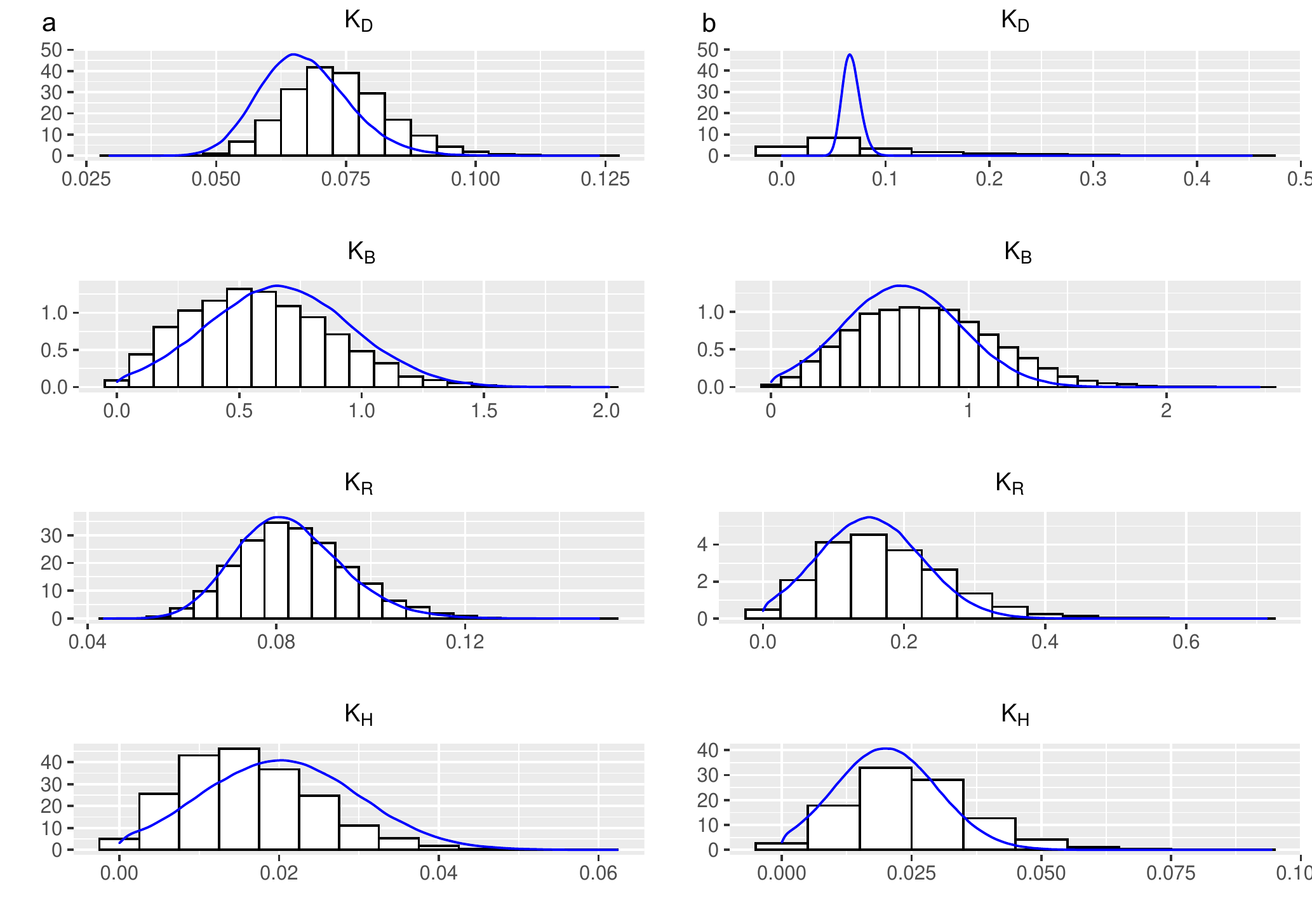}
    \vspace*{-0.5cm}
    \caption{The marginal posterior distributions (histogram) of the decomposition rates of the five-pool BIO-K model in a) Tarlee and b) Brigalow. The blue densities are the prior distributions of the decomposition rates in the five-pool BIO-K model.}
    \label{Pri_Post_5P_BIO-K_Tar_Brig}
\end{figure}

As mentioned earlier in Section \ref{SoilCarbonModel}, prior knowledge of unknown parameters plays a significant role in the presence of small and sparse datasets in a Bayesian setting. We compared a histogram of the samples drawn from the posteriors with the prior distributions of some main model parameters of the three and five-pool BIO-K models to highlight what we have learned about those parameters. Figures \ref{Pri_Post_BIO-K_Tar_Brig}a and \ref{Pri_Post_BIO-K_Tar_Brig}b show the difference between the posterior and prior of the decomposition rate of the SOC and BIO pools of the three-pool BIO-K model in Tarlee and Brigalow, respectively. The difference between the posterior and prior of these parameters of the three-pool BIO-K and regular models of the Broadbalk dataset are shown in Figures \ref{Pri_Post_3P_Roth}a and \ref{Pri_Post_3P_Roth}b, respectively. We presented the difference between the posterior and prior of the decomposition rates in the five-pool BIO-K model of Tarlee and Brigalow in Figures \ref{Pri_Post_5P_BIO-K_Tar_Brig}a and \ref{Pri_Post_5P_BIO-K_Tar_Brig}b, respectively.

We computed the Gelman and Rubin's convergence diagnostics, $\hat{R}$ for the model parameters of the three-pool BIO-K model of the Tarlee, Brigalow, and Broadbalk datasets which are presented in Tables \ref{diag}, \ref{diagBrigalow}, and \ref{diagRoth1}, respectively, in Section \ref{GelmanAndRubin} of the supplementary material. The Gelman and Rubin's convergence diagnostics, $\hat{R}$ for the model parameters of the three-pool Regular model of the Broadbalk dataset are presented in Table \ref{diagRoth2} in Section \ref{GelmanAndRubin} of the supplementary material.

\section{Discussion}\label{SectionConclusion}
In this study, we have developed two SOC models to explore the impact of microbial population growth on estimating the amount of sequestered carbon in farmlands. We have also compared the predictive ability of the three-pool model with the more complex and frequently used five-pool (RothC-like) model. We have implemented these models on three datasets, two of them are small and sparse over time, and the other one is large, to show they are broadly applicable.

Microbial population growth has a positive relationship with carbon decomposition rate and has a dynamic process. The motivating question behind this study is whether considering the decay rates of soil carbon pools as being mediated by the size of the microbial pool in SOC models can improve the accuracy of the SOC models for making inferences on soil carbon dynamics. We fitted the three and five-pool BIO-K models on three datasets, and we found that a three-pool BIO-K model (in all datasets) to have a better predictive ability than the five-pool BIO-K model. Also, we compared the predictive abilities of the three and five-pool BIO-K models with the regular three and five-pool models in \citet{davoudabadi2021modelling} through fitting them on the Tarlee, Brigalow, and Broadbalk datasets. Based on their LFO-CV and WAIC values, the three-pool model introduced by \citet{davoudabadi2021modelling} outperforms the three-pool BIO-K model in both Tarlee and Brigalow but the three-pool BIO-K model has a better predictive ability than the regular three-pool model in the presence of a large dataset, the Broadbalk dataset. Although considering the microbial population growth did not improve the accuracy of the SOC models when making inferences on soil carbon dynamics of small datasets, we showed that the three-pool model outperforms the five-pool model in both regular and BIO-K models. It supports the idea that some concessions in physical realism can lead to better predictive accuracy that can be helpful for national carbon accounting. 

We have successfully shown that a frequently used multi-pool model, the RothC model, might not be as fit-for-purpose compared to the three-pool model when used with both temporally sparse and large datasets. \citet{davoudabadi2021modelling} show that a farmer or a land manager can gain more accuracy in making inferences about soil carbon sequestration when the three-pool model is applied to small datasets. Our study suggests that a three-pool model may not only be suitable for modelling short, short sparse datasets (see \cite{davoudabadi2021modelling}) but also for much longer historical datasets.

One of the advantages of our models and the models introduced in \citet{davoudabadi2021modelling} is that the SOC dynamics are modelled on a yearly time step that makes these models easier to fit and analyse (statistically) compared to models that consider the SOC changes on a monthly time step. Our main argument for embracing the yearly time-step arises from the fact that crop yields are often reported on an annual basis.  For models that operate on a monthly time-step, annual yield data must then be dis-aggregated using some sort of model to try and determine how much carbon entered the soil for each month.  Not only is this dis-aggregation a difficult undertaking, but it is a major source of uncertainty in the modelling since these soil carbon inputs are a major driver of the soil carbon dynamics.

Many long-term datasets are available that our models and methods have not been tested over those datasets to estimate the global $CO_2$ emission from soil. We will consider this in the future study.

\section{Acknowledgments}
We would like to thank CSIRO for providing the Tarlee and Brigalow datasets used in this study. MJD was supported by QUT-CSIRO Digital Agriculture Scholarship and a CSIRO Digital Agriculture Top-Up Scholarship. CD was supported by the Australian Research Council. We
gratefully acknowledge the computational resources provided by QUT's High Performance
Computing (HPC) and Research Support Group. We thank the Lawes Agricultural Trust and Rothamsted Research for data from the e-RA database. The Rothamsted Long-term Experiments National Capability (LTE-NC) is supported by the UK BBSRC (Biotechnology and Biological Sciences Research Council, BBS/E/C/000J0300) and the Lawes Agricultural Trust.

\newpage
\clearpage
\bibliographystyle{apalike}
\bibliography{references}

\section*{Supplementary Material}\label{SuppMater}
\appendix

\section{Notation}
The notations related to latent variables $\mathbf{X}$ at time $t$, their corresponding measured values $\mathbf{Y}$ and some model parameters are presented in Table \ref{TabelNotations}. For the sake of notational simplicity we use $i$ to show fields $\{1, 2, 3\}$ in the Tarlee dataset, soil type $\{1, 2, 3\}$ in the Brigalow dataset, and plots $\{2b, 3, 5, 7, 10, 14\}$ in the Broadbalk dataset. In what follows, we simply use field for all field, soil type and plot.

\begin{table}[!h]
\begin{center}
\small\addtolength{\tabcolsep}{-3pt}
\begin{tabular}{|c | c|} 
\hline
 \cellcolor{gray!60} Notation &  \cellcolor{gray!60} Description \\
\hline
$X_{C(t)}^i$ & The mass of SOC (t/ha)\\
\hline
$X_{W(t)}^i$ & The mass of total wheat dry
matter (t/ha)\\
\hline
$X_{S(t)}^i$ & The mass of total sorghum dry
matter (t/ha)\\
\hline
$X_{Str(t)}^i$ & The mass of total straw dry
matter (t/ha)\\
\hline
$X_{G_W(t)}^i$ & The mass of total grain dry matter produced from wheat (t/ha)\\
\hline
$X_{G_S(t)}^i$ & The mass of total grain dry matter produced from sorghum (t/ha)\\
\hline
$X_{G(t)}^i$ & The mass of total grain dry matter (t/ha)\\
\hline
$X_{P(t)}^i$ & The mass of total pasture dry matter (t/ha)\\
\hline
$X_{IOM(t)}^i$ & The mass of IOM (t/ha)\\
\hline
$X_{B(t)}^i$ & The mass of BIO (t/ha)\\
\hline
$X_{D(t)}^i$ & The mass of DPM (t/ha)\\
\hline
$X_{R(t)}^i$ & The mass of resistant plant material (RPM) (t/ha)\\
\hline
$X_{H(t)}^i$ & The mass of HUM (t/ha)\\
\hline
$Y_{TOC(t)}^i$ & The measured value of total SOC (TOC) (t/ha)\\
\hline
$Y_{W(t)}^i$ & The measured value of total wheat dry matter (t/ha)\\
\hline
$Y_{S(t)}^i$ & The measured value of total sorghum dry matter (t/ha)\\
\hline
$Y_{Str(t)}^i$ & The measured value of total straw dry matter (t/ha)\\
\hline
$Y_{G_W(t)}^i$ & The measured value of total wheat grain dry matter (t/ha)\\
\hline
$Y_{G_S(t)}^i$ & The measured value of total sorghum grain dry matter (t/ha)\\
\hline
$Y_{G(t)}^i$ & The measured value of total grain dry matter (t/ha)\\
\hline
$Y_{P(t)}^i$ & The measured value of total pasture dry matter (t/ha)\\
\hline
$Y_{IOM(t)}^i$ & The measured value of IOM (t/ha)\\
\hline
$Y_{H(t)}^i$ & The measured value of HUM (t/ha)\\
\hline
$Y_{POC(t)}^i$ & The measured value of particulate organic carbon (POC) (t/ha)\\
\hline
$K_C$ & The decay rate of total SOC ($Y^{-1}$)\\
\hline
$K_A$ & The decay rate of the carbon in pool $A$ ($Y^{-1}$)\\
\hline
$\pi_{AB}$ &  Proportion of the mass of carbon transfer \\ & from carbon pool $A$ to carbon pool $B$ \\
\hline
$\Delta t$ & The yearly time step \\
\hline
$P_D$ & Proportion of the carbon input that added to the DPM pool \\
\hline
\end{tabular}
\end{center}
\caption{The notations of latent variables, their corresponding measured values and some model parameters.}
\label{TabelNotations}
\end{table}

All processes and all observations, for instance in Tarlee dataset, at time $t$ in all fields are denoted by $X_{(t)} = (X_{(t)}^1, X_{(t)}^2, X_{(t)}^3)$ and $Y_{(t)} = (Y_{(t)}^1, Y_{(t)}^2, Y_{(t)}^3)$, respectively. All processes at all fields and all times are represented by $\mathbf{X}$, and $\mathbf{Y}$ represents all available data. We denote a set of variables as $Y_{1:t} = (Y_{(1)},...,Y_{(t)})$. The log-normal distribution is denoted by $LN(\mu_1,\sigma_1^2)$ with mean parameter $\mu_1$ and variance parameter $\sigma_1^2$ for a log transformation of the random variable. $N(\mu_2,\sigma_2^2)$ represents the normal distribution with mean and variance $\mu_2$ and $\sigma_2^2$, respectively. Some other notations are presented wherever they are required.

\section{Datasets}\label{datasets}
Tables \ref{ManagementTreatments}, \ref{BrigalowDataTable}, and \ref{BroadbalkDataTable} show the duration of management treatments in Tarlee, Brigalow, and Broadbalk, respectively.

\begin{table}[ht]
\begin{center}
\begin{tabular}{|c|c|c|c|}
\hline
 \cellcolor{gray!60} Management treatments   & \cellcolor{gray!60} Field 1 & \cellcolor{gray!60} Field 2 & \cellcolor{gray!60} Field 3 \\
\hline    
    Wheat for grain & (1979 - 1987) and  & - & - \\ & (1990 - 1996) &  & \\ 
\hline     
    Wheat for hay & 1988 and 1989 & 1989 & - \\
\hline    
    Fallow & 1997 & 1997 & 1997 \\
\hline    
    Wheat for grain and fallow & - & (1979 - 1988) and & - \\ &  &  (1990 - 1996) & \\
\hline 
    Wheat and pasture & - & - & (1979 - 1987) \\
\hline 
    Wheat and pasture for hay & - & - & 1988 and 1989 \\
\hline 
    Wheat for grain and pasture & - & - & (1990 - 1996) \\
\hline

\end{tabular}
\caption{\label{ManagementTreatments} The duration of management treatments in three fields in Tarlee.}
\end{center}
\end{table}

\begin{table}[ht]
\begin{center}
\begin{tabular}{|c|c|c|c|}
\hline
 \cellcolor{gray!60} Management treatments   & \cellcolor{gray!60} Soil type 1 & \cellcolor{gray!60} Soil type 2 & \cellcolor{gray!60} Soil type 3 \\
\hline    
    Cleared & 1982  & 1982 & 1982 \\ 
\hline 
    Wheat for grain & (1985 - 1992) and  & (1985 - 1992) & (1985 - 1992) \\ & (1994, 1996, 1998) & (1994, 1996, 1998) & (1994, 1996, 1998) \\ 
\hline     
    Sorghum for grain & 1984, 1995, & 1984, 1995, & 1984, 1995,  \\ &  1997 and 1999 &  1997 and 1999 & 1997 and 1999 \\
\hline    
    Fallow & 1983 and 1993 & 1983 and 1993 & 1983 and 1993 \\
\hline

\end{tabular}
\caption{\label{BrigalowDataTable} The duration of management treatments in Brigalow.}
\end{center}
\end{table}

\begin{table}[!ht]
\begin{center}
\begin{tabular}{|c|c|}
\hline
 \cellcolor{gray!60} Plot   & \cellcolor{gray!60} Treatment \\
\hline      
    2.2 (2b) &  35 Mg $ha^{-1}$ fresh farmyard manure every year since 1843. \\ 
\hline     
    3 & No fertilizer or organic amendments. \\
\hline    
    5 &  Mineral fertilizer: 35 kg P $ha^{-1}$ 90 kg K $ha^{-1}$ Na, and Mg. No organic amendments, \\ & except that straw was chopped
and returned to the plots for 11 years (1869-1879).\\
\hline    
    7 & N2 PKNaMg (until 1967), N2 PK(Na)Mg (from 1968), N2 PKMg (from 1985),\\ & and N2 (P)KMg (from 2001) \\
\hline    
    10 &  N2, and N4 (from 2001)\\
\hline    
    14 & N2 P $Mg^*$ (until 1967), N2 PK $Mg^*$ (from 1968),  N2 $PKMg^*$ (from 1985),\\ & and N4 $PK^*(Mg^*)$ (from 2001)  \\
\hline
\end{tabular}
\caption{\label{BroadbalkDataTable} The plots and treatments in Broadbalk.}
\end{center}
\end{table}

\section{Prior and Proposal Distributions}\label{PriorAndProposalDists}
The model parameters and their prior probability density functions related to the Tarlee, Brigalow, and Broadbalk datasets are listed in Tables \ref{TabelPrior_LogModel}, \ref{BrigalowTablePrior_LogModel}, and \ref{RothTablePrior_LogModel} (to avoid repetition, the priors of the model parameters which have the same distribution in all datasets are presented in Table \ref{TabelPrior_LogModel}).

\begin{table}[!ht]
\begin{center}
\small\addtolength{\tabcolsep}{-3pt}
\begin{tabular}{|c c c|} 
\hline
 \cellcolor{gray!60} Parameter &  \cellcolor{gray!60} Prior &  \cellcolor{gray!60} Type \\
\hline
$X_{C(1978)}^1$ & Truncated-normal$(40,5^2,lower=0)$ & Uninformative\\
\hline
$X_{C(1978)}^2$ & Truncated-normal$(40,5^2,lower=0)$ & Uninformative\\
\hline
$X_{C(1978)}^3$ & Truncated-normal$(40,5^2,lower=0)$ & Uninformative\\
\hline
$X_{IOM}$ & Truncated-normal$(4,0.5^2,lower=0)$ & Uninformative\\
\hline
$K_C$ & LN$(-2.71,(0.127)^2)$ & Informative \\
\hline
$K_D$ & LN$(-2.71,(0.127)^2)$ & Informative
\\
\hline
$K_B$ &Truncated-normal$(0.66,0.3^2,lower=0)$ & Informative 
\\
\hline
$K_R$ &LN$(-2.5,(0.135)^2)$ & Informative
\\
\hline
$K_H$ &Truncated-normal$(0.02,0.01^2,lower=0)$ & Informative
\\
\hline
$c$ & N$(0.45,(0.01)^2)$ & Informative \\
\hline
$r_W$ & N$(0.5,(0.067)^2)$ & Informative \\
\hline
$r_P$ & N$(1,(0.125)^2)$ & Informative \\
\hline
$p$ & Beta$(89.9,809.1)$ & Informative \\
\hline
$h_W$ & LN$(0.825,(0.36)^2)$ & Weakly Informative \\
\hline
$\mu _{G_W}$ & N$(0.42,(1.18)^2)$ & Weakly Informative \\
\hline
$\mu _P$ & N$(1.41,(1.81)^2)$ & Weakly Informative \\
\hline
$\rho _{G_W}$ & Uniform$(-1,1)$ & Uninformative \\
\hline
$\rho _P$ & Uniform$(-1,1)$ & Uninformative \\
\hline
$\sigma _{\eta}^2$ & Truncated-normal$(0,0.5^2,lower=0)$ & Uninformative \\
\hline
$\sigma _{\eta C}^2$ & Truncated-normal$(0,0.5^2,lower=0)$ & Uninformative \\
\hline
$\sigma _{\eta D}^2$ & Truncated-normal$(0,0.5^2,lower=0)$ & Uninformative \\
\hline
$\sigma _{\eta B}^2$ & Truncated-normal$(0,0.5^2,lower=0)$ & Uninformative \\
\hline
$\sigma _{\eta R}^2$ & Truncated-normal$(0,0.5^2,lower=0)$ & Uninformative \\
\hline
$\sigma _{\eta H}^2$ & Truncated-normal$(0,0.5^2,lower=0)$ & Uninformative \\
\hline
$\sigma _{G_W}^2$ & Truncated-normal$(0,0.5^2,lower=0)$ & Uninformative \\
\hline
$\sigma _{W}^2$ & Truncated-normal$(0,0.5^2,lower=0)$ & Uninformative \\
\hline
$\sigma _{P}^2$ & Truncated-normal$(0,0.5^2,lower=0)$ & Uninformative \\
\hline
$\pi _{DH}$ & Uniform$(0,1)$ & Uninformative \\
\hline
$\pi _{RH}$ & Uniform$(0,1)$ & Uninformative \\
\hline
$\pi _{HH}$ & Uniform$(0,1)$ & Uninformative \\
\hline
$\pi _{BH}$ & Uniform$(0,1)$ & Uninformative \\
\hline
$\pi _{DB}$ & Uniform$(0,1)$ & Uninformative \\
\hline
$\pi _{RB}$ & Uniform$(0,1)$ & Uninformative \\
\hline
$\pi _{HB}$ & Uniform$(0,1)$ & Uninformative \\
\hline
$\pi _{BB}$ & Uniform$(0,1)$ & Uninformative \\
\hline
$\pi _{CB}$ & Uniform$(0,1)$ & Uninformative \\
\hline
$\pi _{BC}$ & Uniform$(0,1)$ & Uninformative \\
\hline
$P_D$ & Uniform$(0,1)$ & Uninformative \\
\hline
$\sigma _{\epsilon TOC}^2$ & 0.025 & Fixed \\
\hline
$\sigma _{\epsilon POC}^2$ & 0.9 & Fixed \\
\hline
$\sigma _{\epsilon G_W}^2$ & 0.023 & Fixed \\
\hline
$\sigma _{\epsilon W}^2$ & 0.133 & Fixed \\
\hline
$\sigma _{\epsilon P}^2$ & 0.067 & Fixed \\
\hline
$\sigma _{\epsilon IOM}^2$ & 0.01 & Fixed \\
\hline
$\sigma _{\epsilon H}^2$ & 0.1 & Fixed \\
\hline
\end{tabular}
\end{center}
\caption{Prior distributions of parameters of the Tarlee dataset and the ones are common in all datasets.}
\label{TabelPrior_LogModel}
\end{table}

\begin{table}[!ht]
\begin{center}
\small\addtolength{\tabcolsep}{-3pt}
\begin{tabular}{|c c c|} 
\hline
 \cellcolor{gray!60} Parameter &  \cellcolor{gray!60} Prior &  \cellcolor{gray!60} Type \\
\hline
$X_{C(1981)}^1$ & Truncated-normal$(60,5^2,lower=0)$ & Uninformative\\
\hline
$X_{C(1981)}^2$ & Truncated-normal$(60,5^2,lower=0)$ & Uninformative\\
\hline
$X_{C(1981)}^3$ & Truncated-normal$(60,5^2,lower=0)$ & Uninformative\\
\hline
$X_{IOM}$ & Truncated-normal$(12,2^2,lower=0)$ & Uninformative\\
\hline
$h_S$ & LN$(0.46, (1.6)^2)$ & Informative \\
\hline
$\rho _{G_S}$ & Uniform$(-1,1)$ & Uninformative \\
\hline
$\mu _{G_S}$ & N$(0.42,(1.18)^2)$ & Weakly Informative \\
\hline
$r_S$ & N$(0.5,(0.067)^2)$ & Informative \\
\hline
$K_C$ & LN$(-2.71,(0.127)^2)$ & Informative \\
\hline
$K_D$ & Truncated-normal$(10,5^2,lower=0)$ & Informative
\\
\hline
$K_R$ &Truncated-normal$(0.15,0.075^2,lower=0)$ & Informative \\
\hline
$\sigma _{G_S}^2$ & Truncated-normal$(0,0.5^2,lower=0)$ & Uninformative \\
\hline
$\sigma _{\eta B}^2$ & Truncated-normal$(0,0.5^2,lower=0)$ & Weakly Informative\\
\hline
$\sigma _{\eta R}^2$ & Truncated-normal$(0,0.5^2,lower=0)$ & Weakly Informative\\
\hline
$\sigma _{\eta H}^2$ & Truncated-normal$(0,0.5^2,lower=0)$ & Weakly Informative\\
\hline
$\sigma _{S}^2$ & Inv-gamma$(0.01,0.01)$ & Uninformative \\
\hline
$\sigma _{\epsilon G_S}^2$ & 0.023 & Fixed \\
\hline
$\sigma _{\epsilon S}^2$ & 0.133 & Fixed \\
\hline
\end{tabular}
\end{center}
\caption{Prior distributions of parameters of the Brigalow dataset.}
\label{BrigalowTablePrior_LogModel}
\end{table}

\begin{table}[!ht]
\begin{center}
\small\addtolength{\tabcolsep}{-3pt}
\begin{tabular}{|c c c|} 
\hline
 \cellcolor{gray!60} Parameter &  \cellcolor{gray!60} Prior &  \cellcolor{gray!60} Type \\
\hline
$X_{IOM}$ & Truncated-normal$(17,2^2,lower=0)$ & Uninformative\\
\hline
$\rho _{G}$ & Uniform$(-1,1)$ & Uninformative \\
\hline
$\rho _{Str}$ & Uniform$(-1,1)$ & Uninformative \\
\hline
$\mu _{G}$ & N$(0.42,(1.18)^2)$ & Weakly Informative \\
\hline
$\mu _{Str}$ & N$(0.42,(1.18)^2)$ & Weakly Informative \\
\hline
$K_C$ & LN$(-2.71,(0.127)^2)$ & Informative \\
\hline
$K_D$ & Truncated-normal$(10,5^2,lower=0)$ & Informative
\\
\hline
$K_R$ &Truncated-normal$(0.15,0.075^2,lower=0)$ & Informative \\
\hline
$\sigma _{G}^2$ & Truncated-normal$(0,0.5^2,lower=0)$ & Uninformative \\
\hline
$\sigma _{Str}^2$ & Truncated-normal$(0,0.5^2,lower=0)$ & Weakly Informative \\
\hline
$\sigma _{\epsilon G}^2$ & 0.023 & Fixed \\
\hline
$\sigma _{\epsilon Str}^2$ & 0.067 & Fixed \\
\hline
$\sigma _{\epsilon H}^2$ & 0.1 & Fixed \\
\hline
$\sigma _{\epsilon POC}^2$ & 0.9 & Fixed \\
\hline
$X_{C(1852)}^{\text{each plot}}$ & 28.8 & Fixed\\
\hline
\end{tabular}
\end{center}
\caption{Prior distributions of parameters of the Broadbalk dataset.}
\label{RothTablePrior_LogModel}
\end{table}

\newpage
\section{Methods}
\subsection{Kalman Filter} \label{Sub.KF}
As the model is a combination of linear and non-linear sub-models, we use the Kalman filter algorithm (Algorithm \ref{euclidKF1}) to draw samples from the posterior and compute the log-likelihood of the linear sub-models. This method is an  optimal estimator in the sense of minimising the variance of the estimated state in the case of linear-Gaussian state-space model which has the following form: 
\begin{align*}
    &X_{(t)}= \boldsymbol{A}^* X_{(t-1)}+ \boldsymbol{B}^* u_{(t)} + \epsilon_{(t)}^* \\
    &Y_{(t)}= \boldsymbol{C}^* X_{(t)} + \nu_{(t)}^* ;
\end{align*}
where $\epsilon_{(t)}^* \sim N(\boldsymbol{0},\boldsymbol{Q}^*)$, $\nu_{(t)}^* \sim N(\boldsymbol{0},\boldsymbol{R}^*)$, $\boldsymbol{A}^*$ is the state-transition matrix, the control-input matrix $\boldsymbol{B}^*$ is applied to a known vector of inputs $u_{(t)}$, and $\boldsymbol{C}^*$  is the observation matrix. Here, the multivariate normal density with mean vector $\boldsymbol{\mu}$ and covariance matrix $\boldsymbol{\Sigma}$ is shown by $\mbox{MVN}(\boldsymbol{\mu}, \boldsymbol{\Sigma})$. In Algorithm \ref{euclidKF1}, the Kalman gain matrix is denoted by $\boldsymbol{K}_{(t)}$, and $X_{(t)}^{t-1}$ and $\boldsymbol{P}_{(t)}^{t-1}$ are the expectations of state variable and the process noise, respectively given all observations up to and including time $t-1$.

\begin{algorithm}
\caption{Kalman filter algorithm}\label{euclidKF1}
\begin{algorithmic}[1]

\State Initialize with initial state $\Hat{X}_{(0)} = x_{(0)}$ and $\Hat{\boldsymbol{P}}_{(0)} = \boldsymbol{Q}^*$ at $t=0$;
\For {$t = 1,...,\textit{T}$}
\State $X_{(t)}^{t-1} = \boldsymbol{A}^* \Hat{X}_{(t-1)} + \boldsymbol{B}^* u_{(t)}$,  \quad \text{State estimate extrapolation};
\State $\boldsymbol{P}_{(t)}^{t-1} = \boldsymbol{A}^* \Hat{\boldsymbol{P}}_{(t-1)}\boldsymbol{A}^{*'}  + \boldsymbol{Q}^*$,  \quad \text{State covariance extrapolation};
\State $\boldsymbol{K}_{(t)} = \boldsymbol{P}_{(t)}^{t-1} \boldsymbol{C}^{*'}[\boldsymbol{R}^* + \boldsymbol{C}^* \boldsymbol{P}_{(t)}^{t-1} \boldsymbol{C}^{*'}]^{-1}$,  \quad \text{Kalman gain matrix};
\State $\Hat{X}_{(t)} = X_{(t)}^{t-1} + \boldsymbol{K}_{(t)} [Y_{(t)} - \boldsymbol{C}^* X_{(t)}^{t-1}]$, \quad \text{State estimate update};
\State $\Hat{\boldsymbol{P}}_{(t)} = [\boldsymbol{I} - \boldsymbol{K}_{(t)} \boldsymbol{C}^*] \boldsymbol{P}_{(t)}^{t-1} $,        \quad \text{State covariance update};
\State Compute the log-likelihood contribution, $l_{(t)}^{\mathrm{KF}}$, at time $t$ through the density $\mbox{MVN}(Y_{(t)} - \boldsymbol{C}^* X_{(t)}^{t-1}, \boldsymbol{R}^* + \boldsymbol{C}^* \boldsymbol{P}_{(t)}^{t-1} \boldsymbol{C}^{*'})$;
\EndFor
\State The complete log-likelihood can be calculated as $L^* = \sum_t l_{(t)}^{\mathrm{KF}}$
\end{algorithmic}
\end{algorithm}

\subsection{Bootstrap Particle Filter}\label{Sub.BPF}
We draw samples from the posterior of latent variables and estimate the log-likelihood of the non-linear sub-models through the bootstrap particle filter. The algorithm of the bootstrap particle filter is shown in Algorithm \ref{euclidBF1}.

\begin{algorithm}
\caption{Bootstrap particle filter algorithm}\label{euclidBF1}
\begin{algorithmic}[1]

\For {$k = 1,...,\textit{N}$}
\State $t=1$, \text{draw sample} $X^{k}_{(1)} \sim p(X_{(1)})$;
\EndFor
\For {$t = 2,...,\textit{T}$}

\For {$k = 1,...,\textit{N}$}
\State Draw sample $X_{(t)}^k \sim p(X_{(t)} \vert X^{*k}_{(t-1)})$; 
\State Calculate weights $w_{(t)}^k = p(Y_{(t)} \vert X_{(t)}^k)$;
\EndFor
\State Estimate the log-likelihood component for the $t^{th}$ observation, $\hat{l}_{(t)} = \log \left(\dfrac{\sum_j w_{(t)}^j}{N}\right)$;

\State Normalise weights $W_{(t)}^k = \dfrac{w_{(t)}^k}{\sum_j w_{(t)}^j}$ for $k \in \{1, 2, \dots, N \}$;

\State Resample with replacement $N$ particles $X_{(t)}^k$ based on the normalised importance weights;
\State Estimate the overall log-likelihood  $L^* = \sum_t \hat{l}_{(t)}$.

\EndFor

\end{algorithmic}
\end{algorithm}

\subsection{Correlated Pseudo-marginal Method}\label{subsecCPM}
The process and observation models in our case depend on a set of unknown static parameters $\boldsymbol{\theta}$, we treat the parameters as random variables and utilise Bayesian approach to estimate $\boldsymbol{\theta}$. Since the posterior distribution and the likelihood are intractable in this study, we utilise correlated pseudo-marginal method, one
of the MCMC methods, to generate a sequence of correlated random samples from a probability
distribution. In order to reduce the variance of the ratio of the likelihood estimators $\hat{p}(\mathbf{Y} \vert \mathbf{X}, \boldsymbol{\theta} ^*)$ and $\hat{p}(\mathbf{Y} \vert \mathbf{X}, \boldsymbol{\theta}_{m-1} )$ in the acceptance ratio of the CPM method, one can correlate them through correlating the auxiliary random numbers $U$, used to obtain these estimators. Algorithms \ref{CPMalgorithm} and \ref{BPF_CPM} show the CPM algorithm and the particle filter with a given set of random numbers, respectively.

\begin{algorithm}[ht]
\caption{Correlated pseudo-marginal algorithm}\label{CPMalgorithm}
\begin{algorithmic}[1]
\State Initialise $\boldsymbol{\theta}_0$;
\For {$m = 1,...,\textit{M}^*$}
\State Sample $\boldsymbol{\theta}^{*} \sim Q(.\vert \boldsymbol{\theta}_{m-1})$;
\State Sample $\xi \sim N(\textbf{0}, \boldsymbol{I})$ and set $U^* = \tau U_{m-1} + \sqrt{1-\tau ^2} \xi$;
\State Compute the estimator $\hat{p} (\mathbf{Y} \vert  \boldsymbol{\theta} ^*, U^*)$ using Algorithm \ref{BPF_CPM}
\State Compute the acceptance ratio:
\begin{align*}
r=\frac{\hat{p} (\mathbf{Y} \vert  \boldsymbol{\theta} ^*, U^*) p(\boldsymbol{\theta} ^*)Q(\boldsymbol{\theta}_{m-1} \vert \boldsymbol{\theta}^*)}{\hat{p}(\mathbf{Y} \vert  \boldsymbol{\theta}_{m-1}, U_{m-1} ) p(\boldsymbol{\theta}_{m-1} )Q(\boldsymbol{\theta}^*\vert \Theta_{m-1} )};
\end{align*}

\State Accept $(\boldsymbol{\theta} ^*, U^*)$ with probability $\min (r,1)$ otherwise, output $(\boldsymbol{\theta}_{m-1}, U_{m-1})$
\EndFor
\end{algorithmic}
\end{algorithm}

\begin{algorithm}
\caption{Particle filter with fixed random numbers}\label{BPF_CPM}
\begin{algorithmic}[1]

\State Sample $U_{(j^*)} \sim N(0,1)$ and $V_{(i^*)} \sim N(0,1)$ for all $j^* \in \{1, \dots, TN \}$ and $i^* \in \{ 1, \dots, T \}$;
\State Sample $X_{(1)}^k \sim p(. \vert U_{1:N}, \boldsymbol{\theta})$ for all $k \in \{1, \dots, N\}$;
\For {$t = 1,...,\textit{T-1}$}
\State Sort the collection $\{X_{(t)}^1, \dots, X_{(t)}^N\}$;
\State Compute importance weights $w_{(t)}^k$ and log-likelihoods $\hat{l}_{(t)} = \log \left(\dfrac{\sum_k w_{(t)}^k}{N}\right)$ for $k \in \{1, \dots, N\}$;
\State Sample $X_{(t)}^k$ based on systematic resampling using random values $V_{1:T}$ and normalised 
\newline \phantom{aa} weights $W_{(t)}^k$ for $k \in \{1, \dots, N\}$;
\State Set $X_{(t+1)}^k$ as a sample from $p(. \vert X_{(t)}^k, U_{Nt+1:N(t + 1)}, \boldsymbol{\theta})$ for $k \in \{1, \dots, N\}$;

\EndFor
\State Estimate the overall log-likelihood  $L^* = \sum_t \hat{l}_{(t)}$.

\end{algorithmic}
\end{algorithm}

\subsection{Predictive Density Estimation}\label{Pred-Density}
We can compute predictive density in the presence of a state-space model as follows
\begin{align*}
    p(Y_{t+1}|Y_{1:t}) = \int _{\Theta} \int _{\mathcal{X}} p(Y_{t+1}|X_t, \theta,Y_{1:t}) p(X_{t}|\theta,Y_{1:t}) p(\theta | Y_{1:t}) d X d \theta.
\end{align*}
As the above integrals typically do not have closed-form solutions, we estimate them based on $J_1$ MCMC samples and $J_2$ particles getting from an SMC algorithm
\begin{align*}
    p(\hat{Y}_{t+1}|Y_{1:t}) = \frac{1}{J_1}\sum _ {j_1=1}^{J_1} \frac{1}{J_2} \sum _{j_2=1}^{J_2} p(Y_{t+1}|X_t^{j_2}, \theta ^{j_1},Y_{1:t}).
\end{align*}
We can estimate the posterior variance of the log predictive density in practice by 
\begin{align*}
    \sum _{t=1}^T \Big\{\frac{1}{S-1}\sum_{s=1}^S [\log p({Y}_{t}| \boldsymbol{\theta}^s) - \overline{\log p({Y}_{t}| \boldsymbol{\theta}^*)} ]^2\Big\}
\end{align*}
where $S = J_1 + J_2$ and $\boldsymbol{\theta}$ includes the vector of parameters $\theta$ and state variables.

\section{Three-pool Model}\label{Three-pool_formula}
\subsection{Process Model}\label{Three_pool_process}
The process model of the three-pool model at time $t$ in field $i$ is:

\begingroup
\allowdisplaybreaks
\begin{align}
\begin{split}
    &\log (X_{C(t)}^i) = \log (X_{C(t-1)}^i e^{-K_C F_{(t-1)} \Delta t} + I_{C(t)}^i \\
    & \qquad \qquad \qquad + \pi _{BC}(U_{(t-1)}^i - \min(U_{(t-1)}^i, \kappa_{BIO}X_{Total(t-1)}^i-X_{B(t-1)}^i)) \\
    & \qquad \qquad \qquad+ X_{B(t-1)}^i(1 - e^{-K_B F_{(t-1)} \Delta t})\pi _{BC}) + \eta _{C(t)}^i, \ \ \ \eta_{C(t)}^i \sim N(0, \sigma_{\eta_C} ^2);
\end{split}
\\
\begin{split}
    &\log (X_{B(t)}^i) = \log (X_{B(t-1)}^i e^{-K_B F_{(t-1)} \Delta t} + \min(U_{(t-1)}^i, \kappa_{BIO}X_{Total(t-1)}^i-X_{B(t-1)}^i))\\ & \qquad \qquad \qquad \qquad \qquad \qquad \qquad \qquad \ \
    + \eta _{B(t)}^i, \ \ \ \eta_{B(t)}^i \sim N(0, \sigma_{\eta_B} ^2);
\end{split}
\\
\begin{split}
& U^i_{(t-1)} = X_{C(t-1)}^i(1 - e^{-K_C F_{(t-1)} \Delta t})\pi _{CB}  + X_{B(t-1)}^i(1 - e^{-K_B F_{(t-1)} \Delta t})\pi _{BB} \nonumber
\end{split}
    \\
    &X_{G_W(t)}^i \sim LN(\mu _{G_W} + \rho _{G_W} (\log (X_{G_W(t-1)}^i)- \mu _{G_W}), \sigma_{G_W}^2);  \label{eq2S}\\
   &X_{W(t)}^i \sim LN(\log h_W + \log (x_{G_W(t)}^i), \sigma_W^2);      \label{eq3S}\\
    &X_{P(t)}^i \sim LN(\mu _P + \rho _P (\log (X_{P(t-1)}^i)- \mu _P), \sigma_P^2); \label{eq4S}\\
    &X_{G_S(t)}^i \sim LN(\mu _{G_S} + \rho _{G_S} (\log (X_{G_S(t-1)}^i)- \mu _{G_S}), \sigma_{G_S}^2); \quad \text{and} \label{eq5sS}\\
    &X_{S(t)}^i \sim LN(\log h_S + \log (x_{G_S(t)}^i), \sigma_S^2); \label{eq6sS}\\ 
    &X_{G(t)}^i \sim LN(\mu _{G} + \rho _{G} (\log (X_{G(t-1)}^i)- \mu _{G}), \sigma_{G}^2);  \label{eq:G_Roth}\\
   & X_{Str(t)}^i \sim LN(\mu _{Str} + \rho _{Str} (\log (X_{Str(t-1)}^i)- \mu _{Str}), \sigma_{Str}^2);\\
    & X_{IOM(t)}^i = M \label{eq:IOM_Roth};
\end{align}
\endgroup
where $M$ is an unknown constant value, $F_{(t-1)} = \dfrac{X_{B(t-1)}^i}{X_{Total(t-1)}^i\kappa_{BIO}}$, and $\kappa_{BIO} = 0.05$. Also, $h_W$, $h_S$, $\rho _P$, $\rho _{G_W}$, $\rho _{G_S}$, and $\rho _{Str}$ denote the harvest index which is the ratio of wheat to grain, the harvest index which is the ratio of sorghum to grain, auto-regressive parameters for the evolution of pasture total dry matter (TDM) and grain TDM of wheat, sorghum, and straw, respectively. The mass of carbon inputs, $I_{C(t)}^i$, for the Tarlee, Brigalow, and Rothamsted datasets is denoted by $IT_{C(t)}^i$, $IB_{C(t)}^i$ and $IR_{C(t)}^i$, respectively which are:

\begin{align*}
    IT_{C(t)}^i = 
\begin{cases}
    c(X_{W(t)}^i-X_{G_W(t)}^i) + c r_W X_{W(t)}^i       & \quad \text{Wheat for Grain }\\
    c p X_{W(t)}^i + c r_W X_{W(t)}^i    & \quad \text{Wheat for Hay }\\
    c X_{P(t)}^i + c r_P X_{P(t)}^i     & \quad \text{Pasture}\\
    c p X_{P(t)}^i + c r_P X_{P(t)}^i   & \quad \text{Pasture for Hay}\\
    0                                & \quad \text{Fallow}
  \end{cases}
\end{align*}

\begin{align*}
    IB_{C(t)}^i = 
\begin{cases}
    c(X_{W(t)}^i-X_{G_W(t)}^i) + c r_W X_{W(t)}^i       & \quad \text{Wheat for Grain }\\
    c p X_{W(t)}^i + c r_W X_{W(t)}^i    & \quad \text{Wheat for Hay }\\
    c(X_{S(t)}^i-X_{G_S(t)}^i) + c r_S X_{S(t)}^i       & \quad \text{Sorghum for Grain }\\
    c p X_{S(t)}^i + c r_S X_{S(t)}^i    & \quad \text{Sorghum for Hay }\\
    0                                & \quad \text{Fallow}
  \end{cases}
\end{align*}
and
\begin{align*}
    IR_{C(t)}^i = 
\begin{cases}
    c X_{Str(t)}^i + c r_W (X_{G(t)}^i + X_{Str(t)}^i)       & \quad \text{Wheat for Grain }\\
    0                                & \quad \text{Fallow}
  \end{cases}
\end{align*}

\noindent where $p$, $r_P$, $r_S$, and $r_W$ are the proportion of the crop left above-ground after harvest, the root-to-shoot ratios (in terms of TDM) for pasture, sorghum and wheat crops, respectively. The amount of carbon that enters into the soil from plant-matters that remains above-ground after harvesting wheat and sorghum for grain are $c(X_{W(t)}^i-X_{G_W(t)}^i)$, and $c(X_{S(t)}^i-X_{G_S(t)}^i)$, respectively. This amount for straw is considered by $c X_{Str(t)}^i$ in the Broadbalk dataset. The amount of carbon comes from below-ground for wheat and sorghum are then $c r_W X_{W(t)}^i$ and $c r_S X_{S(t)}^i$, respectively. Here, $c$ is the carbon content of dry plant matter. The amount of carbon that comes from below and above-ground (i.e., roots and what remains after harvesting) are included in $IT_{C(t)}^i$, $IB_{C(t)}^i$, and $IR_{C(t)}^i$.

\subsection{Observation Model}\label{Three_pool_Obs}
This section is allocated to show the observation model of the three-pool model. The observation models of total organic carbon (TOC), IOM, and input carbon are shown in equations (\ref{3PoolObs})-(\ref{Obs_Str}). As the measurements (\ref{3PoolObs})-(\ref{Obs_Str}) are independent, we can show the joint observation model of them at time $t$ and field $i$ by multiplying the probability of each measurement variable given the corresponding latent variable and parameter $\boldsymbol{\theta}$. The overall observation model across all $i$'s is therefore:
\begin{align*}
    p(Y_{(t)} \vert X_{(t)}, \boldsymbol{\theta}) = \prod _i p(Y_{(t)}^i \vert X_{(t)}^i, \boldsymbol{\theta}),
\end{align*} 
where $Y_{(t)}^i = (Y_{TOC(t)}^i,Y_{IOM(t)}^i, Y_{G_W(t)}^i, Y_{G(t)}^i, Y_{W(t)}^i, Y_{P(t)}^i, Y_{G_S(t)}^i, Y_{Str(t)}^i, Y_{S(t)}^i)$ and \newline $X_{(t)}^i= ( X_{C(t)}^i,X_{B(t)}^i,X_{G_W(t)}^i,X_{G(t)}^i,X_{W(t)}^i,X_{P(t)}^i,X_{G_S(t)}^i,X_{Str(t)}^i,X_{S(t)}^i,X_{IOM(t)}^i)$.

\begin{align}
\begin{split}
    &Y_{TOC(t)}^i \vert X_{C(t)}^i= ~x_{C(t)}^i, X_{IOM(t)}^i= x_{IOM(t)}^i\\
    & \qquad  \qquad \qquad \qquad , X_{B(t)}^i= x_{B(t)}^i \sim LN(\log(x_{C(t)}^i + x_{IOM(t)}^i + x_{B(t)}^i), \sigma_{\epsilon TOC}^2). \label{3PoolObs}
\end{split}
    \\
    &Y_{IOM(t)}^i \vert X_{IOM(t)}^i= x_{IOM(t)}^i \sim LN(\log(x_{IOM(t)}^i), \sigma_{\epsilon IOM}^2); \label{Obs_IOM}\\
    &Y_{G_W(t)}^i \vert X_{G_W(t)}^i= x_{G_W(t)}^i \sim LN(\log(x_{G_W(t)}^i), \sigma_{\epsilon G_W}^2); \label{Obs_G_W}\\
    &Y_{W(t)}^i \vert X_{W(t)}^i= x_{W(t)}^i \sim LN(\log(x_{W(t)}^i), \sigma_{\epsilon W}^2); \label{Obs_W}\\
    &Y_{P(t)}^i \vert X_{P(t)}^i= x_{P(t)}^i \sim LN(\log(x_{P(t)}^i), \sigma_{\epsilon P}^2);  \label{Obs_P}\\
    &Y_{G_S(t)}^i \vert X_{G_S(t)}^i= x_{G_S(t)}^i \sim LN(\log(x_{G_S(t)}^i), \sigma_{\epsilon G_S}^2); \label{Obs_G_S}\\
    &Y_{S(t)}^i \vert X_{S(t)}^i= x_{S(t)}^i \sim LN(\log(x_{S(t)}^i), \sigma_{\epsilon S}^2). \label{Obs_S} \\
    &Y_{G(t)}^i \vert X_{G(t)}^i= x_{G(t)}^i \sim LN(\log(x_{G(t)}^i), \sigma_{\epsilon G}^2); \qquad \qquad \text{and} \label{Obs_G}\\
    &Y_{Str(t)}^i \vert X_{Str(t)}^i= x_{Str(t)}^i \sim LN(\log(x_{Str(t)}^i), \sigma_{\epsilon {Str}}^2).\label{Obs_Str}
\end{align}

\section{Five-pool Model}\label{Five-pool_formula}
\subsection{Process Model}\label{Five_pool_process}
The process model of the five-pool model is shown by (\ref{F_P_M1S})-(\ref{F_P_M4S}). This model also includes sub-models (\ref{eq2S})-(\ref{eq:IOM_Roth}).
\begin{align}
     \log (X_{D(t)}^i) = &\log (X_{D(t-1)}^i e^{-K_D\Delta t} + P_D I_{C(t)}^i) + \eta _{D(t)}^i, \ \ \ \eta_{D(t)}^i \sim N(0, \sigma_{\eta_D} ^2); \label{F_P_M1S}\\
    \log (X_{R(t)}^i) = &\log (X_{R(t-1)}^i e^{-K_R F_{(t-1)}\Delta t} + (1 - P_D) I_{C(t)}^i) + \eta _{R(t)}^i, \ \ \ \eta_{R(t)}^i \sim N(0, \sigma_{\eta_R} ^2);\\
    \log (X_{H(t)}^i) = &\log (X_{H(t-1)}^i e^{-K_H F_{(t-1)} \Delta t} + X_{D(t-1)}^i(1 - e^{-K_D F_{(t-1)} \Delta t})\pi _{DH}\nonumber \\
    & + X_{R(t-1)}^i(1 - e^{-K_R F_{(t-1)} \Delta t})\pi _{RH}  + X_{H(t-1)}^i(1 - e^{-K_H F_{(t-1)} \Delta t})\pi _{HH}\nonumber \\
    & + X_{B(t-1)}^i(1 - e^{-K_B F_{(t-1)} \Delta t})\pi _{BH}) + \pi _{BH} (U^i_{(t-1)} \nonumber \\ 
    & - \min(U^i_{(t-1)}, \kappa_{BIO}X_{Total(t-1)}^i-X_{B(t-1)}^i))  + \eta _{H(t)}^i, \ \ \ \eta_{H(t)}^i \sim N(0, \sigma_{\eta_H} ^2);\\
    \log (X_{B(t)}^i) = &\log (X_{B(t-1)}^i e^{-K_B F_{(t-1)} \Delta t} + \min(U^i_{(t-1)}, \kappa_{BIO}X_{Total(t-1)}^i-X_{B(t-1)}^i)) \sim N(0, \sigma_{\eta_B} ^2); \label{F_P_M4S}
\end{align}

\noindent where 

\begin{align}\label{U_t-1}
    U^i_{(t-1)} = &X_{D(t-1)}^i(1 - e^{-K_D F_{(t-1)} \Delta t})\pi _{DB} + X_{R(t-1)}^i(1 - e^{-K_R F_{(t-1)} \Delta t})\pi _{RB} \nonumber \\
    & + X_{H(t-1)}^i(1 - e^{-K_H F_{(t-1)} \Delta t})\pi _{HB} +  X_{B(t-1)}^i(1 - e^{-K_B F_{(t-1)} \Delta t})\pi _{BB}
\end{align}

Since the transition densities of the latent variables in this model are independent, the transition density of the joint process model of the latent variables can be gained through multiplying them.

\subsection{Observation Model}\label{Five_pool_observation}
The observation models of the IOM pool and carbon input which are presented by equations (\ref{Obs_IOM})-(\ref{Obs_Str}) are the same in this model. The observation models of the TOC, particulate organic carbon (POC), and HUM are  

\begin{align}
\begin{split}
    &\log(Y_{TOC(t)}^i) = \log(x_{D(t)}^i + x_{IOM(t)}^i + x_{B(t)}^i + x_{R(t)}^i + x_{H(t)}^i) +\eta_{\epsilon TOC},\\
    &\qquad  \qquad \qquad \qquad  \qquad  \qquad \qquad \qquad \qquad  \qquad \qquad \qquad \eta_{\epsilon TOC} \sim N(0, \sigma_{\epsilon TOC}^2); \label{5PoolObs1}
\end{split}
     \\
     &\log(Y_{POC(t)}^i) = \log(x_{D(t)}^i + x_{B(t)}^i + x_{R(t)}^i) +\eta_{\epsilon POC}, \ \ \ \ \ \eta_{\epsilon POC} \sim N(0, \sigma_{\epsilon POC}^2); \label{5PoolObs2} \\
     &\log(Y_{H(t)}^i) = \log(x_{H(t)}^i) +\eta_{\epsilon H}, \ \ \ \eta_{\epsilon H} \sim N(0, \sigma_{\epsilon H}^2). \label{5PoolObs3} 
\end{align}

Since the measurements are independent, the joint observation model of them at time $t$ and field $i$, and the overall observation model across all $i$'s can be gained similar to what we mentioned in the observation model of the three-pool model.

\section{Gelman and Rubin’s convergence diagnostic statistic}\label{GelmanAndRubin}

\begin{table}[ht]
\begin{center}
\begin{tabular}{|c|c|c|}
\hline
 \cellcolor{gray!60} Parameter & \cellcolor{gray!60} $\hat{R}$ & \cellcolor{gray!60} Upper C.I. bound on $\hat{R}$ \\
\hline
$K_C$ & 1.00 & 1.00 \\
\hline
$c$ & 1.00 & 1.00 \\
\hline
$r_W$ & 1.00 & 1.00 \\
\hline
$r_P$ & 1.00 & 1.00 \\
\hline
p & 1.00 & 1.00 \\
\hline
$h_W$ & 1.00 & 1.01 \\
\hline
$\mu_{G_W}$ & 1.00 & 1.01 \\
\hline
$\mu_P$ & 1.00 & 1.01 \\
\hline
$\rho_{G_W}$ & 1.00 & 1.01 \\
\hline
$\rho_P$ & 1.00 & 1.01 \\
\hline
$\sigma_{\eta_C}^2$ & 1.00 & 1.01 \\
\hline
$\sigma_{G_W}^2$ & 1.00 & 1.00 \\
\hline
$\sigma_{W}^2$ & 1.01 & 1.02 \\
\hline
$\sigma_{P}^2$ & 1.00 & 1.00 \\
\hline 
$X_{IOM}$ & 1.00 & 1.00 \\
\hline 
$X_{C(1978)}^1$ & 1.00 & 1.00 \\
\hline 
$X_{C(1978)}^2$ & 1.00 & 1.01 \\
\hline 
$X_{C(1978)}^3$ & 1.00 & 1.00 \\
\hline 
$\sigma_{\eta_B}^2$ & 1.01 & 1.03 \\
\hline 
$K_{B}$ & 1.00 & 1.00 \\
\hline 
$\pi_{CB}$ & 1.01 & 1.03 \\
\hline 
$\pi_{BB}$ & 1.00 & 1.00 \\
\hline$
\pi_{BC}$ & 1.00 & 1.00 \\
\hline
\end{tabular}
\caption{\label{diag} The Gelman and Rubin's convergence diagnostic, $\hat{R}$ calculated for model parameters of the three-pool model of the Tarlee dataset. Since the point estimate of $\hat{R}$ for each parameter is less than 1.2, the MCMC samples can be considered to have reached a stationary distribution and are mixing adequately.}
\end{center}
\end{table}

\begin{table}[ht]
\begin{center}
\begin{tabular}{|c|c|c|}
\hline
 \cellcolor{gray!60} Parameter & \cellcolor{gray!60} $\hat{R}$ & \cellcolor{gray!60} Upper C.I. bound on $\hat{R}$ \\
\hline
$K_C$ & 1.00 & 1.00 \\
\hline
$c$ & 1.00 & 1.00 \\
\hline
$r_W$ & 1.00 & 1.00 \\
\hline
p & 1.01 & 1.03 \\
\hline
$h_W$ & 1.00 & 1.00 \\
\hline
$\mu_{G_W}$ & 1.04 & 1.13 \\
\hline
$\mu_{G_S}$ & 1.04 & 1.05 \\
\hline
$\rho_{G_W}$ & 1.00 & 1.01 \\
\hline
$\sigma_{\eta_C}^2$ & 1.01 & 1.02 \\
\hline
$\sigma_{G_W}^2$ & 1.00 & 1.00 \\
\hline
$\sigma_{W}^2$ & 1.00 & 1.00 \\
\hline
$X_{IOM}$ & 1.00 & 1.00 \\
\hline
$X_{C(1982)}^1$ & 1.01 & 1.02 \\
\hline 
$X_{C(1982)}^2$ & 1.01 & 1.02 \\
\hline 
$X_{C(1982)}^3$ & 1.00 & 1.00 \\
\hline 
$r_{S}$ & 1.00 & 1.00 \\
\hline
$\sigma_{\eta_B}^2$ & 1.02 & 1.08 \\
\hline
$\sigma_{S}^2$ & 1.17 & 1.34 \\
\hline
$K_{B}$ & 1.00 & 1.00 \\
\hline
$K_{R}$ & 1.00 & 1.00 \\
\hline
$\pi_{DB}$ & 1.00 & 1.01 \\
\hline
$\pi_{BB}$ & 1.00 & 1.02 \\
\hline
$\pi_{BC}$ & 1.00 & 1.01 \\
\hline
$\pi_{CB}$ & 1.00 & 1.01 \\
\hline
$\mu_{G_S}$ & 1.00 & 1.00 \\
\hline
$\rho_{G_S}$ & 1.00 & 1.00 \\
\hline
$\sigma_{G_S}^2$ & 1.00 & 1.00 \\
\hline
$h_S$ & 1.01 & 1.02 \\
\hline
\end{tabular}
\caption{\label{diagBrigalow} The Gelman and Rubin's convergence diagnostic, $\hat{R}$ calculated for model parameters of the three-pool model of the Brigalow dataset. Since the point estimate of $\hat{R}$ for each parameter is less than 1.2, the MCMC samples can be considered to have reached a stationary distribution and are mixing adequately.}
\end{center}
\end{table}

\begin{table}[ht]
\begin{center}
\begin{tabular}{|c|c|c|}
\hline
 \cellcolor{gray!60} Parameter & \cellcolor{gray!60} $\hat{R}$ & \cellcolor{gray!60} Upper C.I. bound on $\hat{R}$ \\
\hline
$K_C$ & 1.01 & 1.03 \\
\hline
$c$ & 1.00 & 1.00 \\
\hline
$r_W$ & 1.01 & 1.02 \\
\hline
$\mu_{G}$ & 1.03 & 1.10 \\
\hline
$\mu_{Str}$ & 1.00 & 1.00 \\
\hline
$\rho_{G}$ & 1.00 & 1.01 \\
\hline
$\rho_{Str}$ & 1.00 & 1.01 \\
\hline
$\sigma_{\eta_C}^2$ & 1.14 & 1.27 \\
\hline
$\sigma_{G}^2$ & 1.00 & 1.00 \\
\hline
$\sigma_{Str}^2$ & 1.00 & 1.00 \\
\hline
$X_{IOM}$ & 1.00 & 1.01 \\
\hline
$\sigma_{\eta_B}^2$ & 1.06 & 1.08 \\
\hline
$K_{B}$ & 1.00 & 1.00 \\
\hline
$\pi_{BB}$ & 1.00 & 1.00 \\
\hline
$\pi_{BC}$ & 1.01 & 1.03 \\
\hline
$\pi_{CB}$ & 1.00 & 1.00 \\
\hline
\end{tabular}
\caption{\label{diagRoth1} The Gelman and Rubin's convergence diagnostic, $\hat{R}$ calculated for model parameters of the three-pool BIO-K model of the Broadbalk dataset. Since the point estimate of $\hat{R}$ for each parameter is less than 1.2, the MCMC samples can be considered to have reached a stationary distribution and are mixing adequately.}
\end{center}
\end{table}

\begin{table}[ht]
\begin{center}
\begin{tabular}{|c|c|c|}
\hline
 \cellcolor{gray!60} Parameter & \cellcolor{gray!60} $\hat{R}$ & \cellcolor{gray!60} Upper C.I. bound on $\hat{R}$ \\
\hline
$K_C$ & 1.00 & 1.00 \\
\hline
$c$ & 1.00 & 1.01 \\
\hline
$r_W$ & 1.00 & 1.01 \\
\hline
$\mu_{G}$ & 1.00 & 1.00 \\
\hline
$\mu_{Str}$ & 1.00 & 1.00 \\
\hline
$\rho_{G}$ & 1.00 & 1.00 \\
\hline
$\rho_{Str}$ & 1.00 & 1.00 \\
\hline
$\sigma_{\eta_C}^2$ & 1.00 & 1.00 \\
\hline
$\sigma_{G}^2$ & 1.00 & 1.00 \\
\hline
$\sigma_{Str}^2$ & 1.00 & 1.00 \\
\hline
$X_{IOM}$ & 1.00 & 1.00 \\
\hline
$\sigma_{\eta_B}^2$ & 1.04 & 1.12 \\
\hline
$K_{B}$ & 1.02 & 1.05 \\
\hline
$\pi_{BB}$ & 1.00 & 1.01 \\
\hline
$\pi_{BC}$ & 1.16 & 1.46 \\
\hline
$\pi_{CB}$ & 1.02 & 1.03 \\
\hline
\end{tabular}
\caption{\label{diagRoth2} The Gelman and Rubin's convergence diagnostic, $\hat{R}$ calculated for model parameters of the three-pool regular model of the Broadbalk dataset. Since the point estimate of $\hat{R}$ for each parameter is less than 1.2, the MCMC samples can be considered to have reached a stationary distribution and are mixing adequately.}
\end{center}
\end{table}

\end{document}